%% file: sussp-lecture.tex
\documentclass[10pt]{book}
\usepackage{sussp_TnF} 
\usepackage{graphicx}     
\usepackage{makeidx}
\usepackage{epsfig,amsmath,amssymb,amsfonts,color,axodraw,cite}
\usepackage{rotating}  
\makeindex
\include{trimmarks}
\graphicspath{{./figs/}}

\allowdisplaybreaks

\begin{document}

{
\input paperdef

\input{sussp-lecture_titlepage.tex}

\headings{Higgs and Electroweak Physics} 
{Higgs and Electroweak}
{Sven Heinemeyer}
{Instituto de F\'isica de Cantabria (CSIC), Santander, Spain} 


\section{Introduction}

A major goal of the 
particle physics program at the high energy frontier,
currently being pursued at the Fermilab Tevatron collider
and soon to be taken up by the CERN Large Hadron Collider (LHC),
is to unravel the nature of electroweak symmetry breaking.
While the existence of the massive electroweak gauge bosons ($W^\pm,Z$),
together with the successful description of their behavior by
non-abelian gauge theory, 
requires some form of electroweak symmetry breaking to be present in nature, 
the underlying dynamics is not known yet.
An appealing theoretical suggestion for such dynamics is the Higgs mechanism
\cite{higgs-mechanism}, which 
implies the existence of one or more 
Higgs bosons (depending on the specific model considered).
Therefore, the search for Higgs bosons is a major cornerstone
in the physics programs of past, present and future 
high energy colliders.

Many theoretical models employing the Higgs mechanism in
order to account for electroweak symmetry breaking
have been studied in the literature, of which 
the most popular ones are the Standard Model (SM)~\cite{sm}  
and the Minimal Supersymmetric Standard Model (MSSM)~\cite{mssm}.
Within the SM, the Higgs boson is the last undiscovered particle, whereas the
MSSM has a richer Higgs sector, containing three neutral and two charged
Higgs bosons. 
Among alternative theoretical models beyond the SM and the MSSM,
the most prominent are  
the Two Higgs Doublet Model (THDM)~\cite{thdm}, 
non-minimal supersymmetric extensions of the SM 
(e.g.\ extensions of the MSSM by an extra singlet
superfield \cite{NMSSM-etc}),
little Higgs models~\cite{lhm} and 
models with more than three spatial dimensions~\cite{edm}. 

We will discuss the Higgs boson sector in the SM and the MSSM. This
includes their connection to electroweak precision physics and the
searches for the SM and supersymmetric (SUSY) Higgs bosons at the LHC. While
the LHC will discover a SM Higgs boson and, in case that the MSSM is
realized in nature, almost certainly also one or more SUSY Higgs bosons,
a ``cleaner'' experimental environment, such as at the ILC, will be
needed to measure all the Higgs boson characteristics~\cite{lhcilc,lhc2fc}.


\section{The SM and the Higgs}


\subsection{Higgs: Why and How?}

We start with looking at one of the most simple Lagrangians, the one of
QED:
\begin{align}
\cL_{\rm QED} &= -\ed{4} F_{\mu\nu} F^{\mu\nu} 
                 + \bar\psi (i \ga^\mu D_\mu - m) \psi~.
\end{align}
Here $D_\mu$ denotes the covariant derivative
\begin{align}
D_\mu &= \partial_\mu + i\,e\,A_\mu~.
\end{align}
$\psi$ is the electron spinor, and $A_\mu$ is the photon vector
field. The QED Lagrangian is invariant under the local $U(1)$ gauge symmetry, 
\begin{align}
\psi &\to e^{-i\al(x)}\psi~, \\
A_\mu &\to A_\mu + \ed{e} \partial_\mu \al(x)~.
\label{gaugeA}
\end{align}
Introducing a mass term for the photon, 
\begin{align}
\cL_{\rm photon~mass} &= \edz m_A^2 A_\mu A^\mu~,
\end{align}
however, is not gauge-invariant. Applying \refeq{gaugeA} yields
\begin{align}
\edz m_A^2 A_\mu A^\mu &\to \edz m_A^2 \KKL
A_\mu A^\mu + \frac{2}{e} A^\mu \partial_\mu \al 
+ \ed{e^2} \partial_\mu \al \, \partial^\mu \al \KKR~.
\end{align}

A way out is the Higgs mechanism~\cite{higgs-mechanism}. 
The simplest implementation uses one
elementary complex scalar Higgs field~$\Phi$ that has a vacuum
expectation value~$v$ (vev) that is constant in space and time.
The Lagrangian of the new Higgs field reads
\begin{align}
\cL_\Phi &= \cL_{\Phi, {\rm kin}} + \cL_{\Phi, {\rm pot}}
\end{align}
with
\begin{align}
\cL_{\Phi, {\rm kin}} &= (D_\mu \Phi)^* \, (D^\mu \Phi)~, \\
-\cL_{\Phi, {\rm pot}} &= V(\Phi) = \mu^2 |\Phi|^2 + \la |\Phi|^4~.
\end{align}
Here $\la$ has to be chosen positive to have a potential bounded from
below. $\mu^2$ can be either positive or negative, where we will see
that $\mu^2 < 0$ yields the desired vev, as will be shown below.
The complex scalar field $\Phi$ can be parametrized by two real scalar
fields~$\phi$ and~$\eta$, 
\begin{align}
\Phi(x) &= \ed{\wz} \phi(x) e^{i \eta(x)}~,
\end{align}
yielding
\begin{align}
V(\phi) &= \frac{\mu^2}{2} \phi^2 + \frac{\la}{4} \phi^4~.
\end{align}
Minimizing the potential one finds
\begin{align}
\frac{{\rm d}V}{{\rm d}\phi}_{\big| \phi = \phi_0} &=
\mu^2 \phi_0 + \la \phi_0^3 \stackrel{!}{=} 0~.
\end{align}
Only for $\mu^2 < 0$ this yields the desired non-trivial solution
\begin{align}
\phi_0 &= \sqrt{\frac{-\mu^2}{\la}} \KL = \langle \phi \rangle =: v \KR~.
\end{align}
The picture simplifies more by going to the ``unitary gauge'', 
$\al(x) = -\eta(x)/v$, which yields a real-valued $\Phi$ everywhere. 
The kinetic term now reads
\begin{align}
(D_\mu \Phi)^* \, (D^\mu \Phi) &\to 
\edz (\partial_\mu \phi)^2 + \edz e^2 q^2 \phi^2 A_\mu A^\mu~,
\label{LphiA}
\end{align}
where $q$ is the charge of the Higgs field, which can now be expanded around
its vev,
\begin{align}
\phi(x) &= v \; + \; H(x)~.
\label{vH}
\end{align}
The remaining degree of freedom, $H(x)$ is a real scalar boson, the
Higgs boson. The Higgs boson mass and self-interactions are obtained by
inserting \refeq{vH} into the Lagrangian (neglecting a constant term), 
\begin{align}
-\cL_{\rm Higgs} &= \edz \mH^2 H^2 + \frac{\kappa}{3!} H^3
                  + \frac{\xi}{4!} H^4~,
\end{align}
with
\begin{align}
\mH^2 = 2 \la v^2, \quad
\kappa = 3 \frac{\mH^2}{v}, \quad
\xi = 3 \frac{\mH^2}{v^2}~.
\end{align}
Similarly, \refeq{vH} can be inserted in \refeq{LphiA}, yielding (neglecting
the kinetic term for $\phi$), 
\begin{align}
\cL_{\rm Higgs-photon} &= \edz m_A^2 A_\mu A^\mu + e^2 q^2 v H A_\mu A^\mu
+ \edz e^2 q^2 H^2 A_\mu A^\mu
\end{align}
where the second and third term describe the interaction between the
photon and one or two Higgs bosons, respectively, and the first term is
the photon mass, 
\begin{align}
\mA^2 &= e^2 q^2 v^2~.
\label{mA}
\end{align}
Another important feature can be observed: the coupling of the photon to
the Higgs is proportional to its own mass squared.

Similarly a gauge invariant Lagrangian can be defined to give mass to
the chiral fermion $\psi = (\psi_L, \psi_R)^T$,
\begin{align}
\cL_{\rm fermion~mass} &= y_\psi \psi_L^\dagger \, \Phi \, \psi_R + {\rm c.c.}~,
\end{align}
where $y_\psi$ denotes the dimensionless Yukawa coupling. Inserting 
$\Phi(x) = (v + H(x))/\wz$ one finds
\begin{align}
\cL_{\rm fermion~mass} &= m_\psi \psi_L^\dagger \psi_R 
                     + \frac{m_\psi}{v} H\, \psi_L^\dagger \psi_R 
                     + {\rm c.c.}~,
\end{align}
with 
\begin{align}
m_\psi &= y_\psi \frac{v}{\wz}~.
\end{align}
Again the important feature can be observed: by construction the
coupling of the fermion to the Higgs boson is proportional to its own
mass $m_\psi$.

The ``creation'' of a mass term can be viewed from a different
angle. The interaction of the gauge field or the fermion field with the
scalar background field, i.e.\ the vev, shift the masses of these fields
from zero to non-zero values. This is shown graphically in
\reffi{fig:masses} for the gauge boson (a) and the fermion (b) field.
%
\setlength{\unitlength}{0.25mm}
\vspace{3em}
\begin{figure}[htb!]
\begin{center}
\begin{picture}(60,10)(80,40)
\Photon(0,25)(50,25){3}{6}
\LongArrow(55,25)(70,25)
\put(-15,30){$V$}
\put(-15,50){$(a)$}
\end{picture}
\begin{picture}(60,10)(40,40)
\Photon(0,25)(50,25){3}{6}
\put(75,30){$+$}
\end{picture}
\begin{picture}(60,10)(15,40)
\Photon(0,25)(50,25){3}{6}
\DashLine(25,25)(12,50){3}
\DashLine(25,25)(38,50){3}
\Line(9,53)(15,47)
\Line(9,47)(15,53)
\Line(35,53)(41,47)
\Line(35,47)(41,53)
\put(15,80){$v$}
\put(50,80){$v$}
\put(75,30){$+$}
\end{picture}
\begin{picture}(60,10)(-10,40)
\Photon(0,25)(75,25){3}{9}
\DashLine(20,25)(8,50){3}
\DashLine(20,25)(32,50){3}
\DashLine(55,25)(43,50){3}
\DashLine(55,25)(67,50){3}
\Line(5,53)(11,47)
\Line(5,47)(11,53)
\Line(29,53)(35,47)
\Line(29,47)(35,53)
\Line(40,53)(46,47)
\Line(40,47)(46,53)
\Line(64,53)(70,47)
\Line(64,47)(70,53)
\put(08,80){$v$}
\put(43,80){$v$}
\put(57,80){$v$}
\put(92,80){$v$}
\put(110,30){$+ \cdots$}
\end{picture} \\
\begin{picture}(60,80)(80,20)
\ArrowLine(0,25)(50,25)
\LongArrow(55,25)(70,25)
\put(-15,32){$f$}
\put(-15,50){$(b)$}
\end{picture}
\begin{picture}(60,80)(40,20)
\ArrowLine(0,25)(50,25)
\put(75,30){$+$}
\end{picture}
\begin{picture}(60,80)(15,20)
\ArrowLine(0,25)(25,25)
\ArrowLine(25,25)(50,25)
\DashLine(25,25)(25,50){3}
\Line(22,53)(28,47)
\Line(22,47)(28,53)
\put(43,70){$v$}
\put(75,30){$+$}
\end{picture}
\begin{picture}(60,80)(-10,20)
\ArrowLine(0,25)(25,25)
\ArrowLine(25,25)(50,25)
\ArrowLine(50,25)(75,25)
\DashLine(25,25)(25,50){3}
\DashLine(50,25)(50,50){3}
\Line(22,53)(28,47)
\Line(22,47)(28,53)
\Line(47,53)(53,47)
\Line(47,47)(53,53)
\put(43,70){$v$}
\put(78,70){$v$}
\put(110,30){$+ \cdots$}
\end{picture}  \\
\end{center}
\caption{%
Generation of a gauge boson mass (a) and a fermion mass (b) via the
interaction with the vev of the Higgs field.
}
\label{fig:masses}
\end{figure}
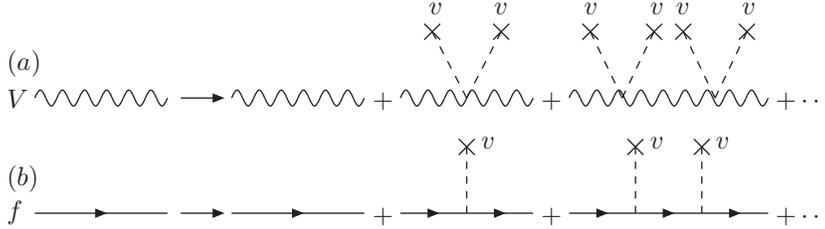

\noindent
The shift in the propagators reads (with $p$ being the external momentum
and $g = e q$ in \refeq{mA}):
\begin{align}
&(a) \; &\ed{p^2} \; \to \; \ed{p^2} + 
    \sum_{k=1}^{\infty} \ed{p^2} \KKL \KL \frac{g v}{2} \KR \ed{p^2} \KKR^k
    &= \ed{p^2 - m_V^2} 
    {\rm ~with~} m_V^2 = g^2 \frac{v^2}{4} ~, \\
&(b) \; &\ed{\pslash} \; \to \; \ed{\pslash} + 
    \sum_{k=1}^{\infty} \ed{\pslash} \KKL \KL \frac{y_\psi v}{2} \KR 
                                        \ed{\pslash} \KKR^k
    &= \ed{\pslash - m_\psi} 
    {\rm ~with~} m_\psi = y_\psi \frac{v}{\wz} ~.
\end{align}


\subsection{SM Higgs Theory}

We now turn to the electroweak sector of the SM, which is described by
the gauge symmetry $SU(2)_L \times U(1)_Y$. The bosonic part of the
Lagrangian is given by
\begin{align}
\cL_{\rm bos} &= -\ed{4} B_{\mu\nu} B^{\mu\nu} 
- \ed{4} W_{\mu\nu}^a W^{\mu\nu}_a
+ |D_\mu \Phi|^2 - V(\Phi), \\
V(\Phi) &= \mu^2 |\Phi|^2 + \la |\Phi|^4~.
\end{align}
$\Phi$ is a complex scalar doublet with charges $(2, 1)$ under the SM
gauge groups, 
\begin{align}
\Phi &= \VL \phi^+ \\ \phi^0 \VR~,
\end{align}
and the electric charge is given by $Q = T^3 + \edz Y$,
where $T^3$ the third component of the weak isospin. We furthermore have
\begin{align}
D_\mu &= \partial_\mu + i g \frac{\tau^a}{2} W_{\mu\,a} 
                     + i g^\prime \frac{Y}{2} B_\mu ~, \\
B_{\mu\nu} &= \partial_\mu B_\nu - \partial_\nu B_\mu ~, \\
W_{\mu\nu}^a &= \partial_\mu W_\nu^a - \partial_\nu W_\mu^a 
               - g f^{abc} W_{\mu\,b} W_{\nu\,c}~.
\end{align}
$g$ and $g^\prime$ are the $SU(2)_L$ and $U(1)_Y$ gauge couplings,
respectively, $\tau^a$ are the Pauli matrices, and $f^{abc}$ are the
$SU(2)$ structure constants.

Choosing $\mu^2 < 0$ the minimum of the Higgs potential is found at
\begin{align}
\langle \Phi \rangle &= \ed{\wz} \VL 0 \\ v \VR 
\quad {\rm with} \quad 
v:= \sqrt{\frac{-\mu^2}{\la}}~.
\end{align}
$\Phi(x)$ can now be expressed through the vev, the Higgs boson and
three Goldstone bosons $\phi_{1,2,3}$, 
\begin{align}
\Phi(x) &= \ed{\wz} \VL \phi_1(x) + i \phi_2(x) \\ 
                        v + H(x) + i \phi_3(x) \VR~.
\end{align}
Diagonalizing the mass matrices of the gauge bosons, one finds that
the three massless Goldstone bosons are absorbed as longitudinal
components of the three massive gauge bosons, $W_\mu^\pm, Z_\mu$, while the
photon $A_\mu$ remains massless, 
\begin{align}
W_\mu^\pm &= \ed{\wz} \KL W_\mu^1 \mp i W_\mu^2 \KR ~,\\
Z_\mu &= \cw W_\mu^3 - \sw B_\mu ~,\\
A_\mu &= \sw W_\mu^3 + \cw B_\mu ~.
\end{align}
Here we have introduced the weak mixing angle 
$\theta_W = \arctan(g^\prime/g)$, and $\sw := \sin \theta_W$, 
$\cw := \cos \theta_W$. The Higgs-gauge boson interaction Lagrangian
reads, 
\begin{align}
\cL_{\rm Higgs-gauge} &= \KKL \MW^2 W_\mu^+ W^{-\,\mu} 
                           + \edz \MZ^2 Z_\mu Z^\mu \KKR 
                      \KL 1 + \frac{H}{v} \KR^2 \non \\
&\quad - \edz \MH^2 H^2 - \frac{\kappa}{3!} H^3 - \frac{\xi}{4!} H^4~,
\end{align}
with 
\begin{align}
\MW &= \edz g v, \quad
\MZ =  \edz \sqrt{g^2 + g^{\prime 2}} \; v, \\
(\MHSM := ) \; \MH &= \sqrt{2 \la}\; v, \quad
\kappa = 3 \frac{\MH^2}{v}, \quad
\xi = 3 \frac{\MH^2}{v^2}~.
\end{align}
From the measurement of the gauge boson masses and couplings one finds
$v \approx 246 \gev$. Furthermore the two massive gauge boson masses are
related via 
\begin{align}
\frac{\MW}{\MZ} &= \frac{g}{\sqrt{g^2 + g^{\prime 2}}} \; = \; \cw~.
\end{align}

We now turn to the fermion masses, where we take the top- and
bottom-quark masses as a representative example. The Higgs-fermion
interaction Lagrangian reads
\begin{align}
\label{SMfmass}
\cL_{\rm Higgs-fermion} &= y_b Q_L^\dagger \, \Phi \, b_R \; + \;
                        y_t Q_L^\dagger \, \Phi_c \, t_R + {\rm ~h.c.}
\end{align}
$Q_L = (t_L, b_L)^T$ is the left-handed $SU(2)_L$ doublet. Going to the
``unitary gauge'' the Higgs field can be expressed as
\begin{align}
\Phi(x) &= \ed{\wz} \VL 0 \\ v + H(x) \VR~,
\label{SMPhi}
\end{align} 
and it is obvious that this doublet can give masses only to the
bottom(-type) fermion(s). A way out is the definition of 
\begin{align}
\Phi_c &= i \si^2 \Phi^* \; = \; \ed{\wz} \VL v + H(x) \\ 0 \VR~,
\label{SMPhic}
\end{align}
which is employed to generate the top(-type) mass(es) in
\refeq{SMfmass}. 
Inserting \refeqs{SMPhi}, (\ref{SMPhic}) into \refeq{SMfmass} yields
\begin{align}
\cL_{\rm Higgs-fermion} &= \mb \bar b b \KL 1 + \frac{H}{v} \KR
                      + \mt \bar t t \KL 1 + \frac{H}{v} \KR
\end{align}
where we have used
$\bar \psi \psi = \psi_L^\dagger \psi_R + \psi_R^\dagger \psi_L$ and
$\mb = y_b v/\wz$, $\mt = y_t v/\wz$.

\bigskip
The mass of the SM Higgs boson, $\MHSM$ is the last remaining free
parameter in the model. However, it is possible to derive bounds on
$\MHSM$ derived from theoretical
considerations~\cite{RGEla1,RGEla2,RGEla3} and from experimental
precision data. Here we review the first approach, while the latter one
is followed in \refse{sec:ewpo}. 

Evaluating loop diagrams as shown in the middle and right of
\reffi{fig:RGEla} yields the renormalization group equation (RGE) for
$\la$, 
\begin{align}
\frac{{\rm d}\la}{{\rm d} t} &=
\frac{3}{8 \pi^2} \KKL \la^2 + \la y_t^2 - y_t^4 
     + \ed{16} \KL 2 g^4 + (g^2 + g^{\prime 2})^2 \KR \KKR~,
\label{RGEla}
\end{align}
with $t = \log(Q^2/v^2)$, where $Q$ is the energy scale.

\setlength{\unitlength}{0.25mm}
\vspace{1.5em}
\begin{figure}[htb!]
\begin{center}
\begin{picture}(90,80)(60,-10)
\DashLine(0,50)(25,25){3}
\DashLine(0,0)(25,25){3}
\DashLine(50,50)(25,25){3}
\DashLine(50,0)(25,25){3}
\put(-15,65){$H$}
\put(-15,-5){$H$}
\put(70,-5){$H$}
\put(70,65){$H$}
\put(30,45){$\la$}
\end{picture}
\begin{picture}(90,80)(10,-10)
\DashLine(0,50)(25,25){3}
\DashLine(0,0)(25,25){3}
\DashLine(75,50)(50,25){3}
\DashLine(75,0)(50,25){3}
\DashCArc(37.5,25)(12.5,0,360){3}
\put(-15,65){$H$}
\put(-15,-5){$H$}
\put(47,57){$H$}
\put(110,-5){$H$}
\put(110,65){$H$}
\end{picture}
\begin{picture}(50,80)(-40,2.5)
\DashLine(0,0)(25,25){3}
\DashLine(0,75)(25,50){3}
\DashLine(50,50)(75,75){3}
\DashLine(50,25)(75,0){3}
\ArrowLine(25,25)(50,25)
\ArrowLine(50,25)(50,50)
\ArrowLine(50,50)(25,50)
\ArrowLine(25,50)(25,25)
\put(-15,100){$H$}
\put(-15,-5){$H$}
\put(53,77){$t$}
\put(110,-5){$H$}
\put(110,100){$H$}
\end{picture}  \\
\caption{%
Diagrams contributing to the evolution of the Higgs self-interaction
$\la$ at the tree level (left) and at the one-loop level (middle and right).
}
\label{fig:RGEla}
\end{center}
\end{figure}
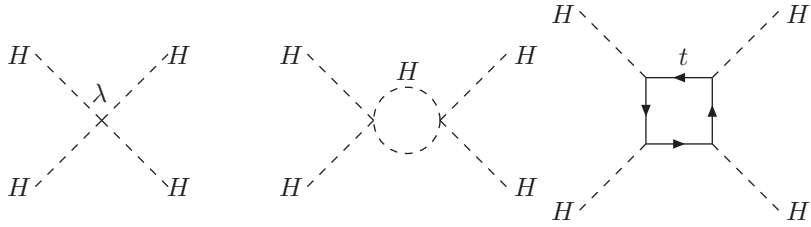

\noindent
For large $\MH^2 \propto \la$ \refeq{RGEla} reduces to
\begin{align}
\frac{{\rm d} \la}{{\rm d} t} &= \frac{3}{8 \pi^2} \la^2 \\
\Rightarrow \quad \la(Q^2) &= \frac{\la(v^2)}
            {1 - \frac{3 \la(v^2)}{8 \pi^2} \log \KL \frac{Q^2}{v^2} \KR}~.
\end{align}
For $\frac{3 \la(v^2)}{8 \pi^2} \log \KL \frac{Q^2}{v^2} \KR = 1$ one
finds that $\la$ diverges (it runs into the ``Landau pole''). 
Requiring $\la(\La) < \infty$
yields an upper bound on $\MH^2$ depending up to which scale $\La$ the
Landau pole should be avoided, 
\begin{align}
\la(\La) < \infty \; \Rightarrow \; 
\MH^2 \le \frac{8 \pi^2 v^2}{3 \log \KL \frac{\La^2}{v^2} \KR}~.
\label{MHup}
\end{align}

\noindent
For small $\MH^2 \propto \la$, on the other hand, \refeq{RGEla} reduces
to
\begin{align}
\frac{{\rm d} \la}{{\rm d} t} &= \frac{3}{8 \pi^2} 
\KKL -y_t^4 + \ed{16} \KL 2 g^4 + (g^2 + g^{\prime 2})^2 \KR \KKR \\
\Rightarrow \quad \la(Q^2) &= \la(v^2) \frac{3}{8 \pi^2}
\KKL -y_t^4 + \ed{16} \KL 2 g^4 + (g^2 + g^{\prime 2})^2 \KR \KKR
\log\KL\frac{Q^2}{v^2}\KR~.
\end{align}
Demanding $V(v) < V(0)$, corresponding to $\la(\La) > 0$ one finds a
lower bound on $\MH^2$ depending on $\La$, 
\begin{align}
\la(\La) > 0 \; \Rightarrow \; 
\MH^2 \; > \; \frac{v^2}{4 \pi^2}
\KKL  -y_t^4 + \ed{16} \KL 2 g^4 + (g^2 + g^{\prime 2})^2 \KR \KKR
\log\KL\frac{\La^2}{v^2}\KR~.
\label{MHlow}
\end{align}

\noindent
The combination of the upper bound in \refeq{MHup} and the lower bound
in \refeq{MHlow} on $\MH$ is shown in \reffi{fig:MHbounds}.
Requiring the validity of the SM up to the GUT scale yields a limit on
the SM Higgs boson mass of $130 \gev \lsim \MHSM \lsim 180 \gev$.

\begin{figure}[htb!]
\vspace{1em}
\includegraphics[height=6cm,angle=90]{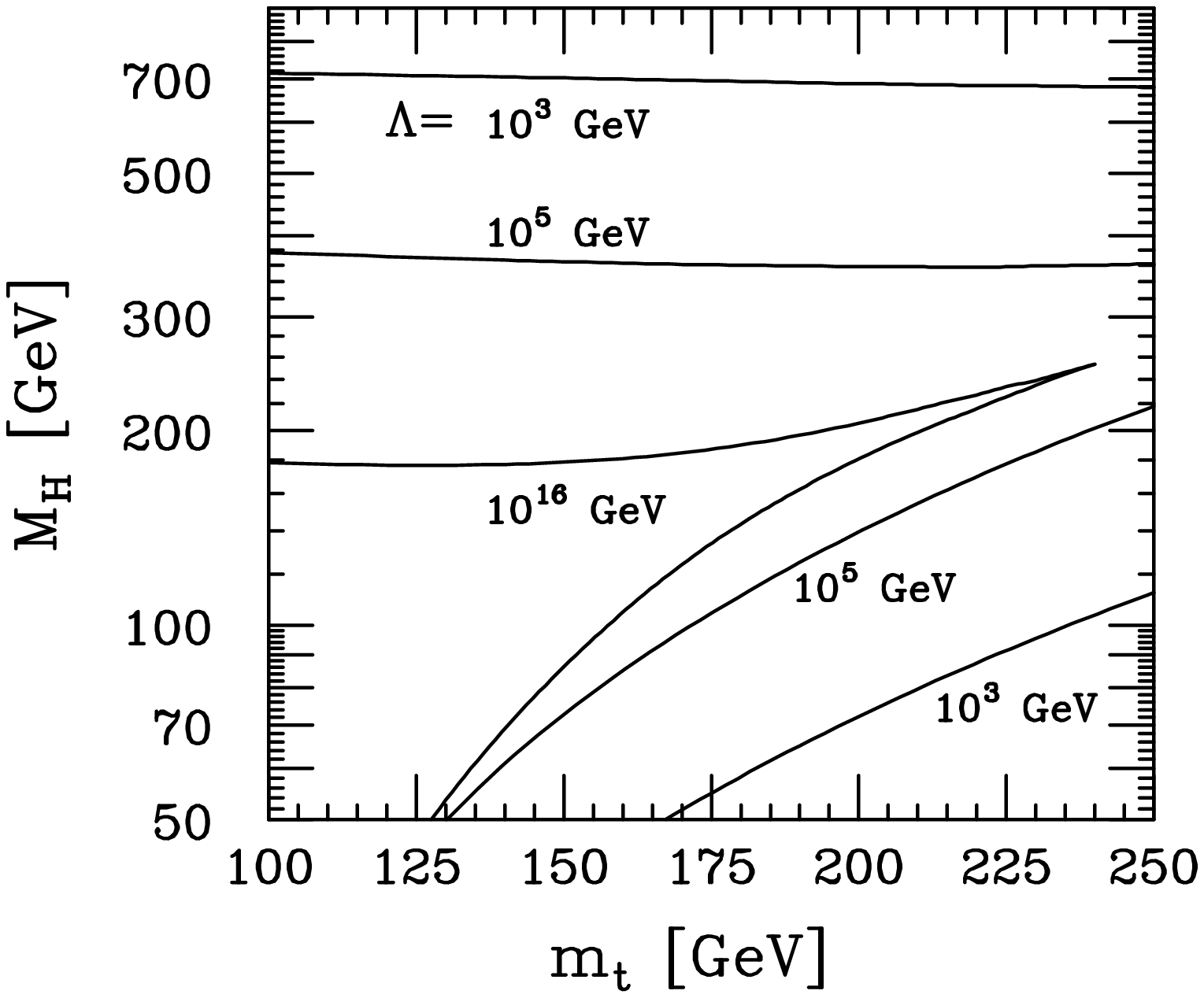}
\includegraphics[height=5cm]{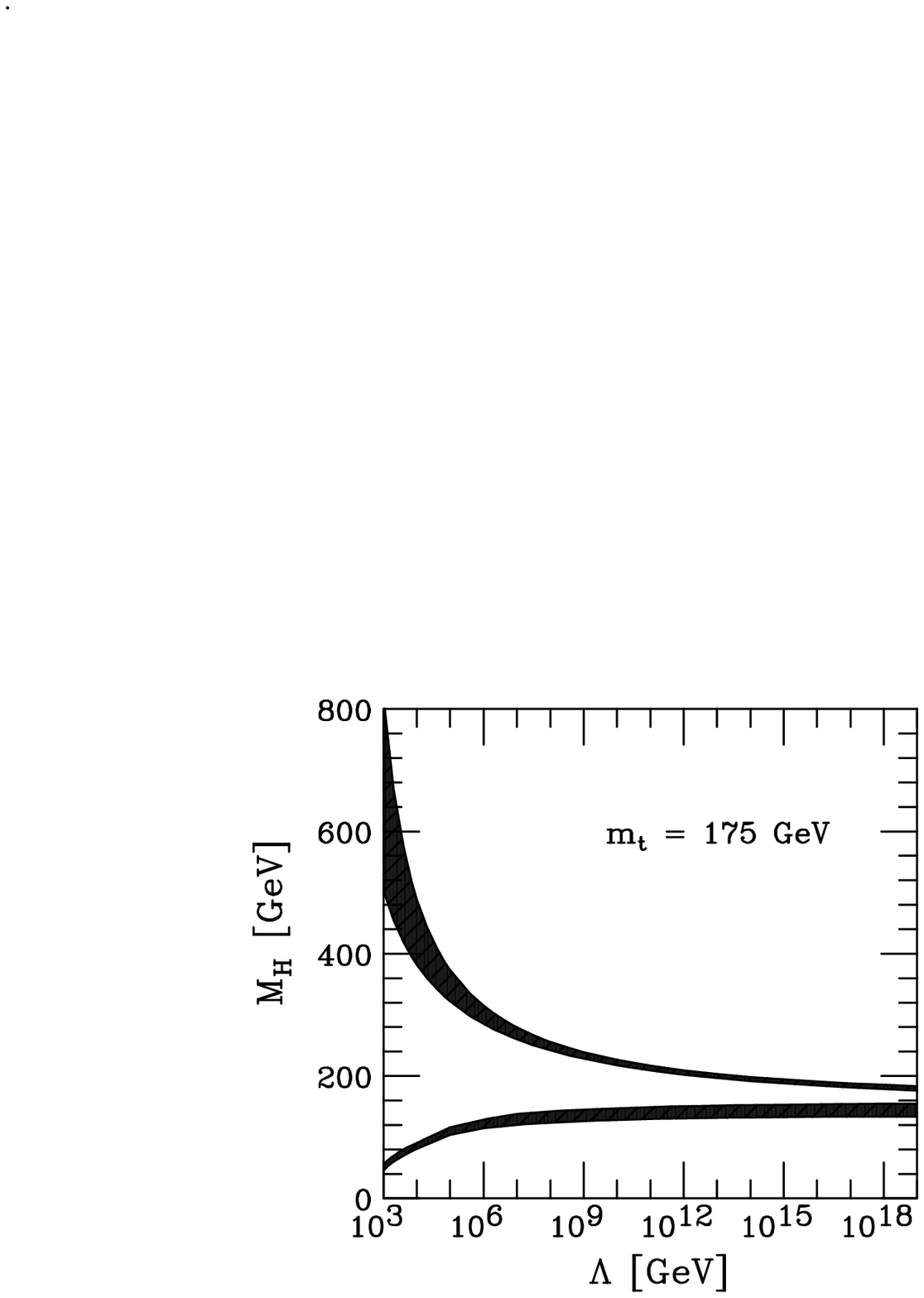}
\caption{%
Bounds on the mass of the Higgs boson in the SM. $\La$ denotes the
energy scale up to which the model is valid~\cite{RGEla1,RGEla2,RGEla3}.
}
\label{fig:MHbounds}
\vspace{-1em}
\end{figure}


\subsection{SM Higgs boson searches at the LHC}
\label{sec:SMHiggsLHC}

A SM-like Higgs boson can be produced in many channels at the LHC as
shown in \reffi{fig:LHCHiggsXS} (taken from \citere{sigmaH}, where also
the relevant original references can be found).
The corresponding discovery potential for a SM-like Higgs boson 
of ATLAS is shown in \reffi{fig:ATLASHiggsdiscovery}~\cite{atlas},
where similar results have been obtained for CMS~\cite{cms}.
With $10\,\ifb$ a $5\,\si$ discovery is expected for 
$\MHSM \gsim 130 \gev$. For lower masses a higher integrated luminosity
will be needed, see also \citere{lhc2fc} for a recent overview.
The largest production cross section is reached by $gg \to H$, which
however, will be visible only in the decay to SM gauge bosons. A precise
mass measurement of $\de\Mh^{\rm exp} \approx 200 \mev$ can be provided by
the decays $H \to \ga\ga$ at lower 
Higgs masses and by $H \to ZZ^{(*)} \to 4\ell$ at higher masses. 
This guarantees the detection of the new state and a precise mass measurement
over the relevant parameter space within the SM.

\begin{figure}[htb!]
\begin{center}
\includegraphics[angle=-90,width=11cm]{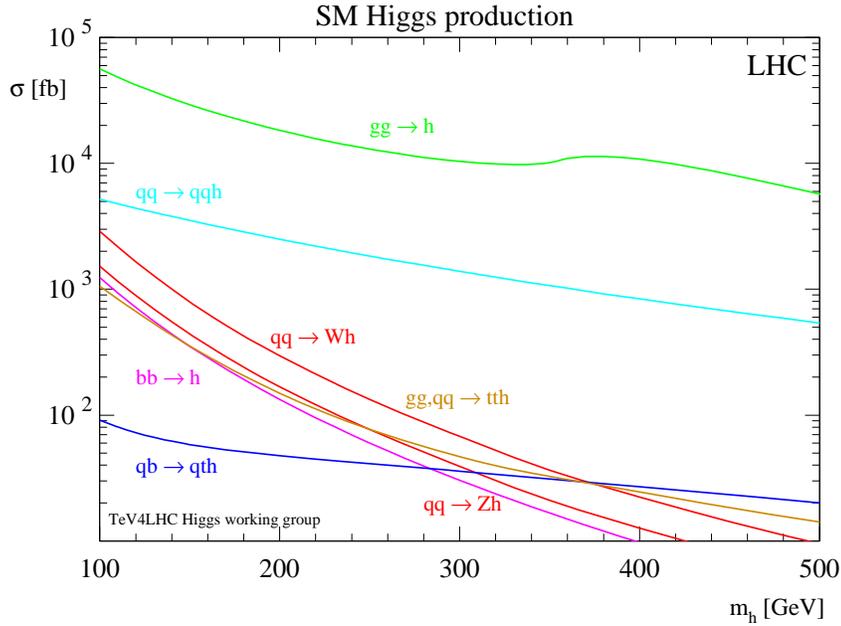}
\caption{%
The various production cross sections for a SM-like Higgs boson at the
LHC are shown as a function of $\MHSM$ (taken from \citere{sigmaH},
where also the relevant references can be found).
}
\label{fig:LHCHiggsXS}
\end{center}
\vspace{-4em}
\end{figure}
%
\begin{figure}[b!]
\begin{center}
\includegraphics[width=12cm]{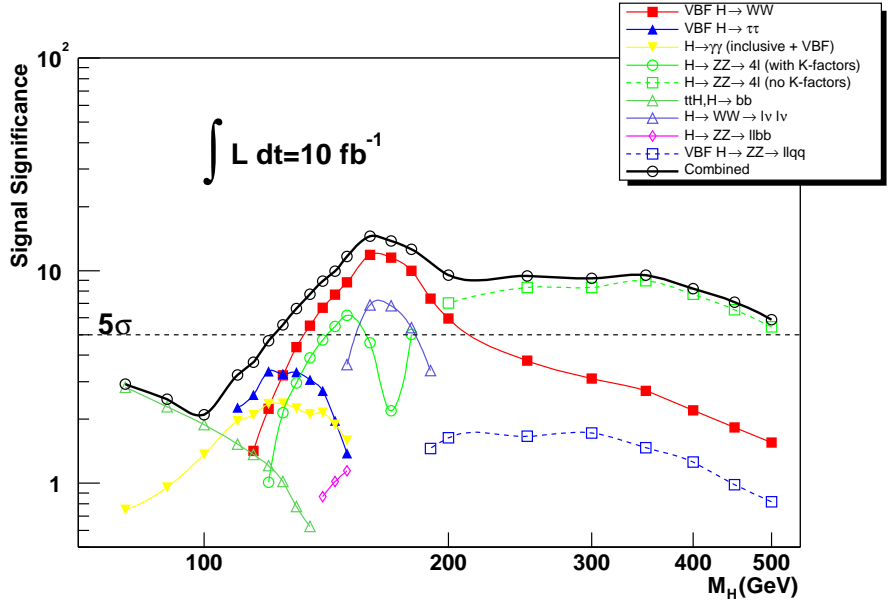}
\caption{%
Significance of a Higgs signal, measured at ATLAS with 
$10\,\ifb$~\cite{atlas}.
Similar results have been obtained for CMS~\cite{cms}.
}
\label{fig:ATLASHiggsdiscovery}
\end{center}
\vspace{-3em}
\end{figure}


\newpage
\subsection{Electroweak precision observables}
\label{sec:ewpo}

Within the SM the electroweak precision observables (EWPO) have been
used to constrain the last unknown 
parameter of the model, the Higgs-boson mass $\MHSM$. 
Originally the EWPO comprise over thousand measurements of ``realistic
observables'' (with partially correlated uncertainties) such as cross
sections, asymmetries, branching ratios etc. This huge set is reduced to
17 so-called ``pseudo observables'' by the LEP~\cite{LEPEWWG} and
Tevatron~\cite{TEVEWWG} Electroweak working groups. 
The ``pseudo observables'' (again called EWPO in the following) comprise
the $W$~boson mass $\MW$, the width of the
$W$~boson, $\Ga_W$, as well as various $Z$~pole observables: the
effective weak mixing angle, $\sweff$, $Z$~decay widths to SM fermions, 
$\Ga(Z \to f \bar f)$, the invisible and total width, $\Ga_{\rm inv}$ and
$\Ga_Z$, forward-backward and left-right asymmetries, $A_{\rm FB}^f$ and
$A_{\rm LR}^f$, and the total hadronic cross section, $\si^0_{\rm had}$.
The $Z$~pole results including their combination are
final~\cite{lepewwg}. Experimental progress from the Tevatron comes for
$\MW$ and $\mt$. (Also the error combination for $\MW$ and $\Ga_W$ from
the four LEP experiments has not been finalized yet due to not-yet-final
analyses on the color-reconnection effects.)

The EWPO that give the strongest constraints on $\MHSM$ are $\MW$, 
$A_{\rm FB}^b$ and $A_{\rm LR}^e$. The value of $\sweff$ is extracted
from a combination of various $A_{\rm FB}^f$ and $A_{\rm LR}^f$, where 
$A_{\rm FB}^b$ and $A_{\rm LR}^e$ give the dominant contribution.

The one-loop contributions to $\MW$
can be decomposed as follows~\cite{sirlin},
\begin{align}
\label{eq:delr}
\MW^2 \KL 1 - \frac{\MW^2}{\MZ^2}\right) &= 
\frac{\pi \al}{\sqrt{2} \GF} \left(1 + \De r\KR , \\
\De r_{1-{\rm loop}} &= \De\al - \frac{\cw^2}{\sw^2}\De\rho 
                     + \De r_{\rm rem}(\MHSM) .
\label{eq:deltar1l}
\end{align}
The first term, $\De\al$ contains large logarithmic contributions as
$\log(\MZ/m_f)$ and amounts $\sim 6\%$. The second term contains the 
$\rho$~parameter~\cite{rho}, being $\De\rho \sim \mt^2$.
This term amounts $\sim 3.3\%$. 
The quantity $\De\rho$, 
\begin{align}
\De\rho &= \frac{\Si^Z(0)}{\MZ^2} - \frac{\Si^W(0)}{\MW^2} ,
\label{delrho}
\end{align}
parameterizes the leading universal corrections to the electroweak
precision observables induced by
the mass splitting between fields in an isospin doublet.
$\Si^{Z,W}(0)$ denote the transverse parts of the 
unrenormalized $Z$ and $W$ boson
self-energies at zero momentum transfer, respectively.
%
The final term in \refeq{eq:deltar1l} is
$\De r_{\rm rem} \sim \log(\MHSM/\MW)$, and with a size of $\sim 1\%$
correction yields the constraints on $\MHSM$. The fact that the leading
correction involving $\MHSM$ is logarithmic also applies to the other
EWPO. Starting from two-loop order, also terms $\sim (\MHSM/\MW)^2$
appear. The SM prediction of $\MW$ as a function of $\mt$ for the range
$\MHSM = 114 \gev \ldots 1000 \gev$ is shown as the dark shaded (green)
band in \reffi{fig:MWMTSM}~\cite{LEPEWWG}. The upper edge with 
$\MHSM = 114 \gev$ corresponds to the lower limit on $\MHSM$ obtained at
LEP~\cite{LEPHiggsSM}. The prediction is compared
with the direct experimental result~\cite{MW80399,mt1731},
\begin{align}
\label{MWexp}
\MW^{\rm exp} &= 80.399 \pm 0.023 \gev ~, \\
\label{mtexp}
\mt^{\rm exp} &= 173.1 \pm 1.3 \gev~,
\end{align}
shown as the dotted (blue) ellipse and with the
indirect results for $\MW$ and $\mt$ as obtained from EWPO (solid/red
ellipse). 
The direct and indirect determination have significant overlap,
representing a non-trivial success for the SM.
However, it should be noted that the experimental value of $\MW$ is
somewhat higher than the region allowed by the LEP Higgs bounds:
$\MHSM \approx 60 \gev$ is preferred as a central value by the
measurement of $\MW$ and $\mt$. 

\begin{figure}[htb!]
\vspace{-1em}
\begin{minipage}[c]{0.5\textwidth}
\includegraphics[width=.99\textwidth]{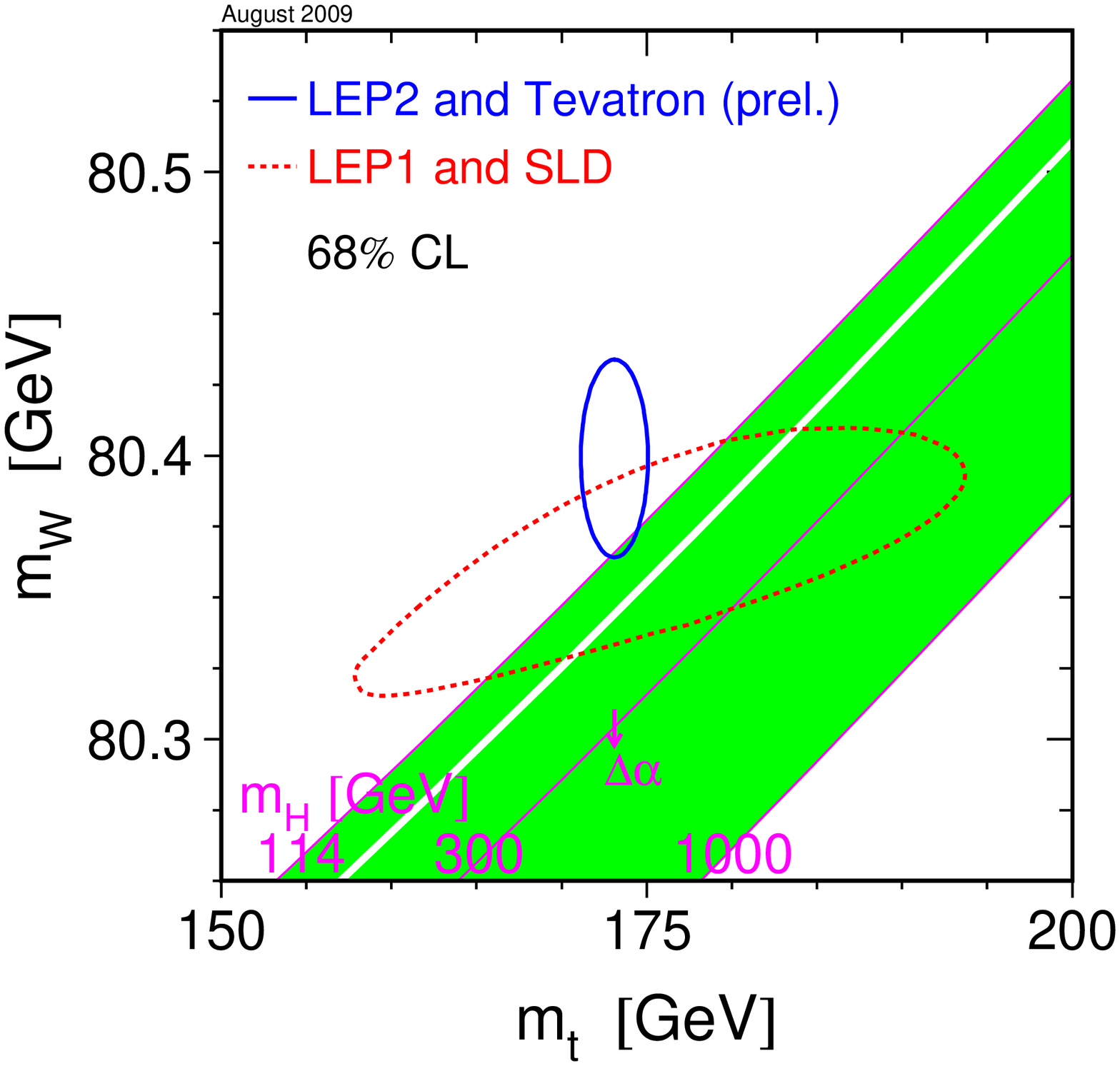}
\end{minipage}
\begin{minipage}[c]{0.03\textwidth}
$\phantom{0}$
\end{minipage}
\begin{minipage}[c]{0.45\textwidth}
\caption{%
Prediction for $\MW$ in the SM as a function of $\mt$ for the range 
$\MHSM = 114 \gev \ldots 1000 \gev$~\cite{LEPEWWG}. The prediction is
compared with  
the present experimental results for $\MW$ and $\mt$ as well as with the
indirect constraints obtained from EWPO.
}
\label{fig:MWMTSM}
\end{minipage}
\vspace{-1em}
\end{figure}

The effective weak mixing angle is evaluated from various asymmetries
and other EWPO as shown in \reffi{fig:sw2effSM}~\cite{gruenewald07}. The
average determination yields $\sweff = 0.23153 \pm 0.00016$ with a
$\chi^2/{\rm d.o.f}$ of $11.8/5$, corresponding to a probability of
$3.7\%$~\cite{gruenewald07}. The large $\chi^2$ is driven by the two
single most precise measurements, $A_{\rm LR}^e$ by SLD and 
$A_{\rm FB}^b$ by LEP, where the earlier (latter) one prefers a 
value of $\MHSM \sim 32 (437) \gev$~\cite{gruenewaldpriv}. 
The two measurements differ by more than $3\,\si$.
The averaged value of $\sweff$, as shown in \reffi{fig:sw2effSM},
prefers $\MHSM \sim 110 \gev$~\cite{gruenewaldpriv}. 

\begin{figure}[htb!]
\vspace{-1em}
\begin{minipage}[c]{0.5\textwidth}
\includegraphics[width=.99\textwidth]{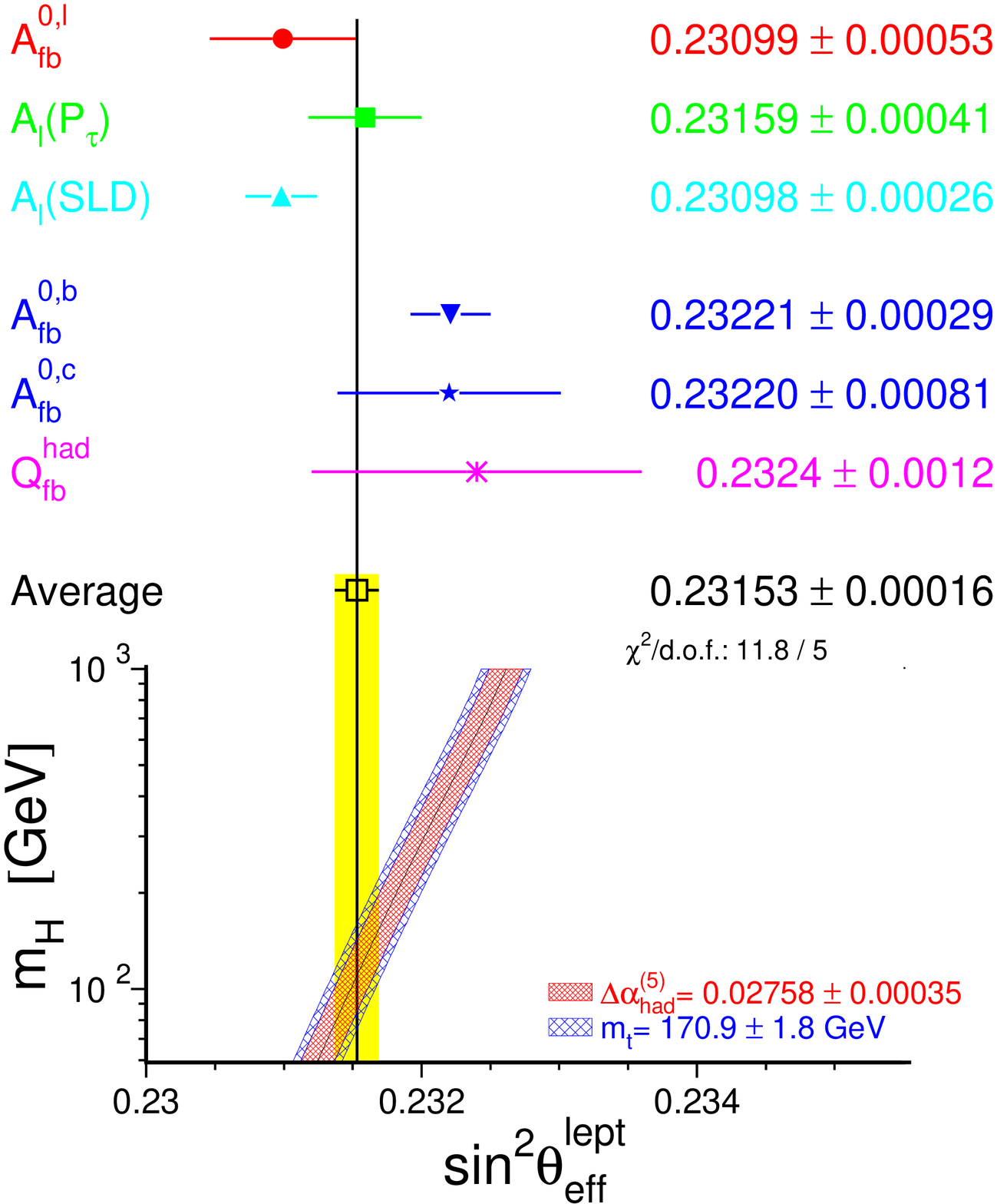}
\end{minipage}
\begin{minipage}[c]{0.03\textwidth}
$\phantom{0}$
\end{minipage}
\begin{minipage}[c]{0.45\textwidth}
\caption{%
Prediction for $\sweff$ in the SM as a function of $\MHSM$ for 
$\mt = 170.9 \pm 1.8 \gev$ and 
$\De\al_{\rm had}^5 = 0.02758 \pm 0.00035$~\cite{gruenewald07}. The
prediction is compared with  
the present experimental results for $\sweff$ as averaged over several
individual measurements.
}
\label{fig:sw2effSM}
\end{minipage}
\vspace{-1em}
\end{figure}

The indirect $\MHSM$ determination for several individual EWPO is given
in \reffi{fig:MHSM}. Shown are the central
values of $\MHSM$ and the one~$\si$ errors~\cite{LEPEWWG}. 
The dark shaded (green) vertical band indicates the combination of the
various single measurements in the $1\,\si$ range. The vertical line shows
the lower LEP bound for $\MHSM$~\cite{LEPHiggsSM}.
It can be seen that $\MW$, $A_{\rm LR}^e$ and $A_{\rm FB}^b$ give the
most precise indirect $\MHSM$ determination, where only the latter one
pulls the preferred $\MHSM$ value up, yielding a averaged value
of~\cite{LEPEWWG} 
\begin{align}
\MHSM = 87^{+35}_{-26} \gev~,
\label{MHSMfit}
\end{align}
still compatible with the direct LEP bound of~\cite{LEPHiggsSM}
\begin{align}
\MHSM \ge 114.4 \gev \mbox{~at~} 95\% \mbox{~C.L.}
\label{MHSMlimit}
\end{align}
Thus, the measurement of $A_{\rm FB}^b$ prevents the SM from being
incompatible with the direct bound and the indirect constraints on
$\MHSM$. 

\begin{figure}[htb!]
\begin{minipage}[c]{0.5\textwidth}
\includegraphics[width=.99\textwidth,height=.99\textwidth]{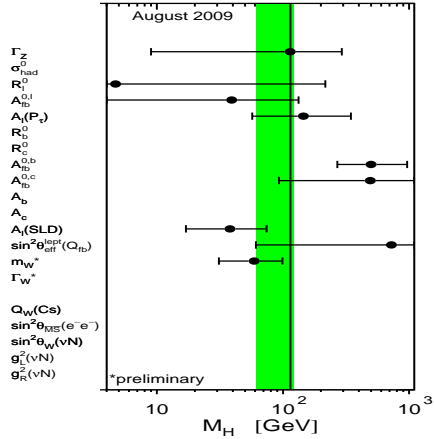}
\end{minipage}
\begin{minipage}[c]{0.03\textwidth}
$\phantom{0}$
\end{minipage}
\begin{minipage}[c]{0.45\textwidth}
\caption{%
Indirect constrains on $\MHSM$ from various EWPO. Shown are the central
values and the one~$\si$ errors~\cite{LEPEWWG}. 
The dark shaded (green) vertical band indicates the combination of the
various single measurements in the $1\,\si$ range. The vertical line shows
the lower bound of $\MHSM \ge 114.4 \gev$ obtained at
LEP~\cite{LEPHiggsSM}.
}
\label{fig:MHSM}
\end{minipage}
\end{figure}

In the left plot of \reffi{fig:blueband}~\cite{LEPEWWG} we show the result for
the global fit to $\MHSM$ including all EWPO, but not including the
direct search bounds from LEP and the Tevatron. $\De\chi^2$ is shown as a
function of $\MHSM$, yielding \refeq{MHSMfit} as best fit with an upper
limit of $157 \gev$ at 95\%~C.L. 
The theory (intrinsic) uncertainty in the SM calculations (as evaluated with 
{\tt TOPAZ0}~\cite{topaz0} and {\tt ZFITTER}~\cite{zfitter}) are
represented by the thickness of the blue band. The width of the parabola
itself, on the other hand, is determined by the experimental precision of
the measurements of the EWPO and the input parameters.
The result changes somewhat if the direct bounds on $\MHSM$ from
LEP~\cite{LEPHiggsSM} and the Tevatron~\cite{TevHiggsSM160170} are taken into
account as shown in the right plot of \reffi{fig:blueband}. The upper
limit reduced to $\MHSM \lsim 150 \gev$ at the 95\%~C.L.~\cite{GFitter}.
In this analysis an Tevatron exclusion of 
$160 \gev < \MHSM < 170 \gev$~\cite{TevHiggsSM160170} was assumed. The most
recent limit is slightly smaller, 
$163 \gev < \MHSM < 166 \gev$~\cite{TevHiggsSM}, however the picture is
expected to vary only very little with this shift.

\begin{figure}[htb!]
\vspace{-1em}
\includegraphics[width=.45\textwidth]{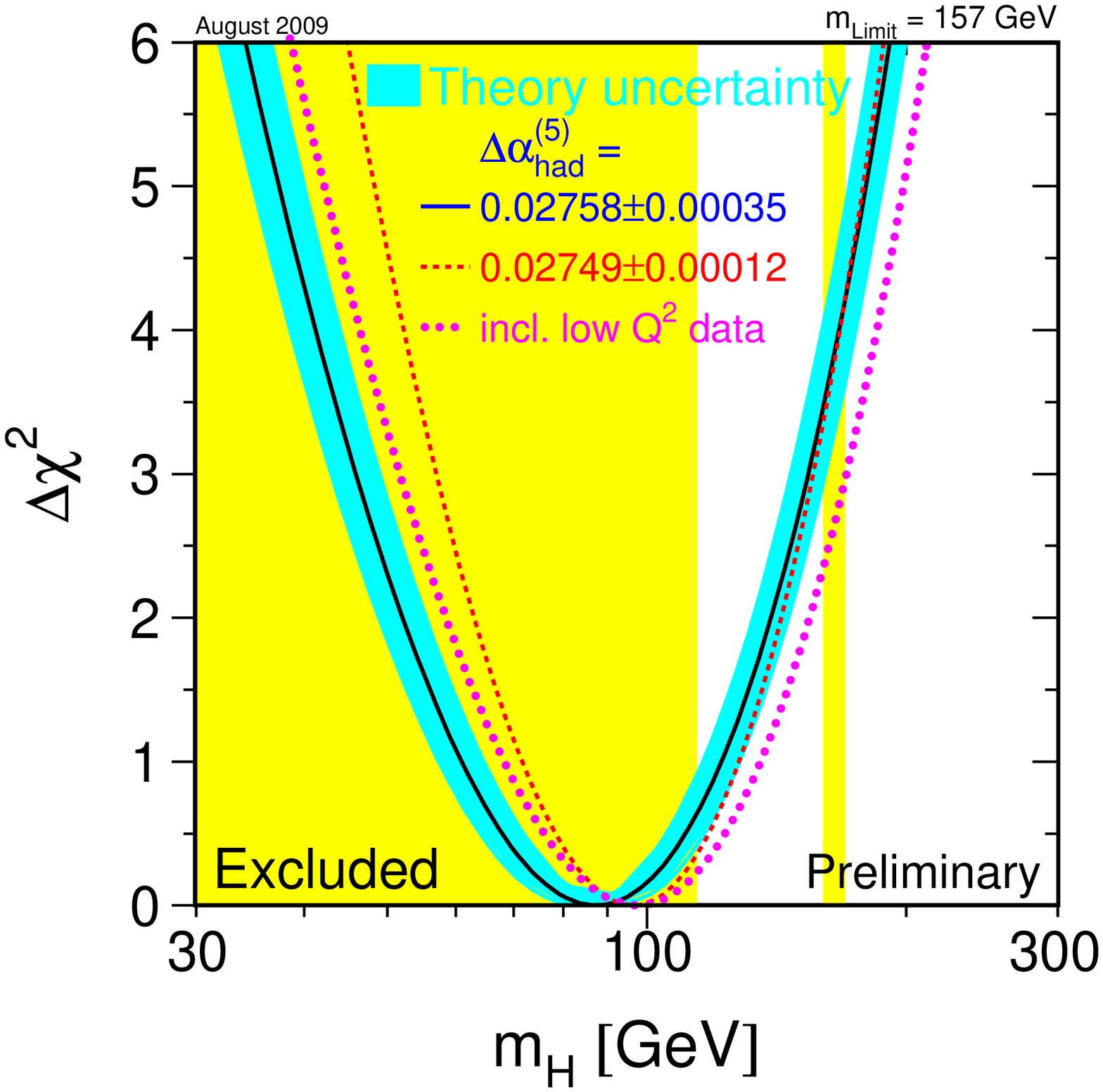}~~
\includegraphics[width=.45\textwidth,height=5.5cm]
                {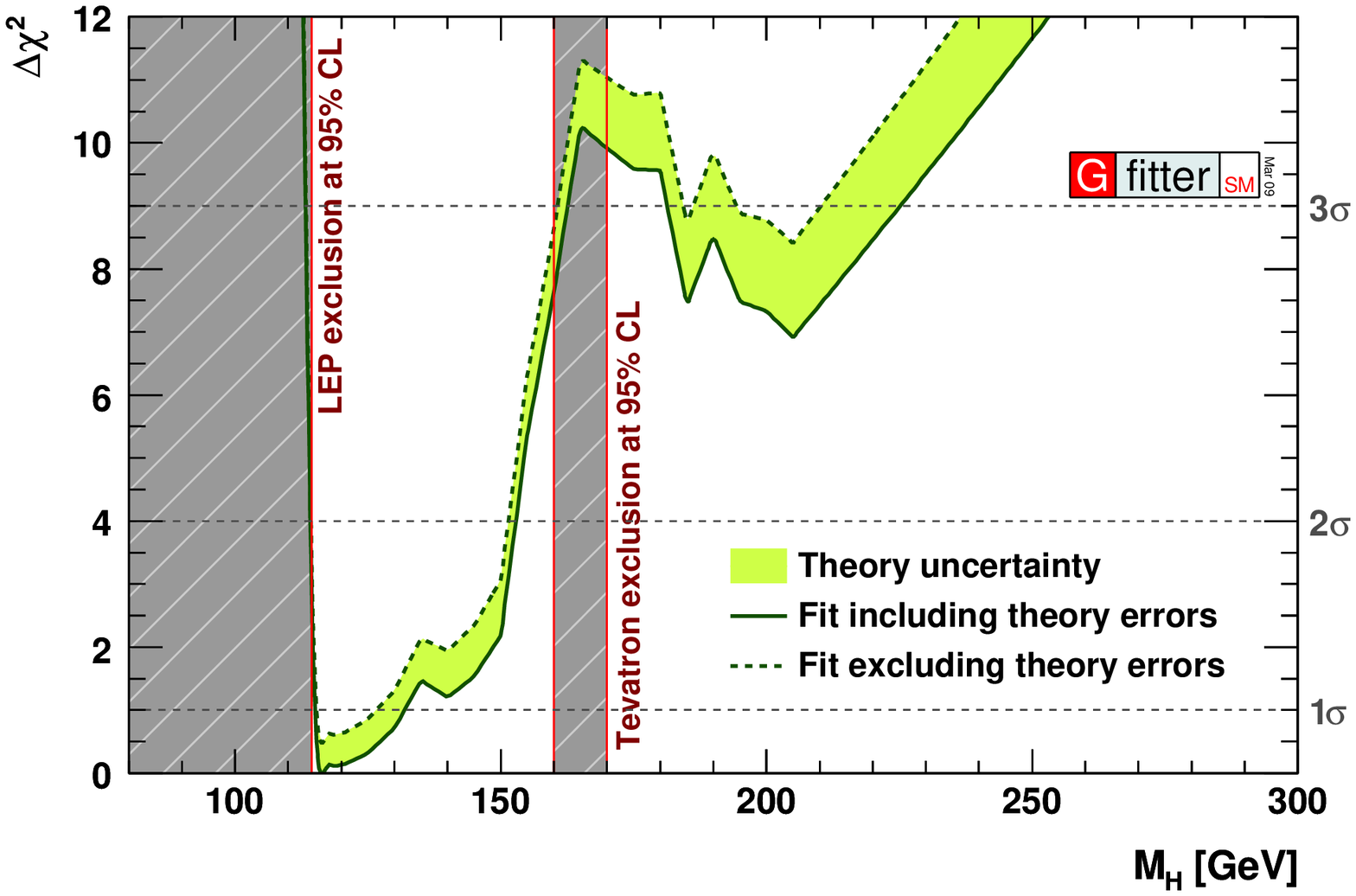}
\caption{%
$\De\chi^2$ curve derived from all EWPO measured at LEP, SLD, CDF and
D0, as a function of $\MHSM$, assuming the SM to be the correct theory
of nature. Left: the direct bounds on $\MHSM$ are not
included~\cite{LEPEWWG}, right: the bounds from LEP~\cite{LEPHiggsSM}
and the Tevatron~\cite{TevHiggsSM} are included~\cite{GFitter}.
}
\label{fig:blueband}
\end{figure}

The current and anticipated future experimental uncertainties for
$\sweff$, $\MW$ and $\mt$ are summarized in \refta{tab:POfuture}. Also
shown is the relative precision of the 
indirect determination of $\MHSM$~\cite{gruenewald07}.
Each column represents the combined results of all detectors and
channels at a given collider, taking into account correlated
systematic uncertainties, see \citeres{blueband,gigaz,moenig,mwgigaz}
for details. The indirect $\MHSM$ determination has to be compared with
the (possible) direct measurement at the LHC~\cite{atlas,cms} and the
ILC~\cite{Snowmass05Higgs}, 
\begin{align}
\label{deltaMHLHC}
\de\MHSM\mbox{}^{\rm ,exp,LHC} &\approx 200 \mev ,\\
\label{deltaMHILC}
\de\MHSM\mbox{}^{\rm ,exp,ILC} &\approx 50 \mev .
\end{align}

\begin{table}[htb!]
\renewcommand{\arraystretch}{1.5}
\begin{center}
\begin{tabular}{|c||c|c|c|c|c|}
\cline{2-6} \multicolumn{1}{c||}{}
& now & Tevatron & LHC & ~ILC~ & ILC with GigaZ \\
\hline\hline
$\de\sweff(\times 10^5)$ & 16   & --- & 14--20 & --- & 1.3\\
\hline
$\de\MW$ [MeV]           & 23   &  20 & 15     & 10  & 7 \\
\hline
$\de\mt$ [GeV]           &  1.3 &  1.0 &  1.0  & 0.2 & 0.1 \\
\hline
$\de\MHSM/\MHSM$ [\%]     &  37 &     &  28    &     & 16\\
\hline
\end{tabular}
\renewcommand{\arraystretch}{1}
\caption{
Current and anticipated future experimental uncertainties for
$\sweff$, $\MW$ and $\mt$. Also shown is the relative precision of the
indirect determination of $\MHSM$~\cite{gruenewald07}.
Each column represents the combined results of all detectors and
channels at a given collider, taking into account correlated
systematic uncertainties, see \citeres{blueband,gigaz,moenig,mwgigaz}
for details. 
}
\label{tab:POfuture}
\end{center}
\end{table}

\noindent
This comparison will shed light on the basic theoretical components for
generating the masses of the fundamental particles.  
On the other hand, an observed inconsistency would be a clear
indication for the existence of a new physics scale.


\newpage
\section{The Higgs in Supersymmetry}

\subsection{Why SUSY?}

Theories based on Supersymmetry (SUSY)~\cite{mssm} are widely
considered as the theoretically most appealing extension of the SM.
They are consistent with the approximate
unification of the gauge coupling constants at the GUT scale and
provide a way to cancel the quadratic divergences in the Higgs sector
hence stabilizing the huge hierarchy between the GUT and the Fermi
scales. Furthermore, in SUSY theories the breaking of the electroweak
symmetry is naturally induced at the Fermi scale, and the lightest
supersymmetric particle can be neutral, weakly interacting and
absolutely stable, providing therefore a natural solution for the dark
matter problem.

The Minimal Supersymmetric Standard Model (MSSM)
constitutes, hence its name, the minimal supersymmetric extension of the
SM. The number of SUSY generators is $N=1$, the smallest possible value.
In order to keep anomaly cancellation, contrary to the SM a second
Higgs doublet is needed~\cite{glawei}.
All SM multiplets, including the two Higgs doublets, are extended to
supersymmetric multiplets, resulting in scalar partners for quarks and
leptons (``squarks'' and ``sleptons'') and fermionic partners for the
SM gauge boson and the Higgs bosons (``gauginos'' and ``gluinos'').   
So far, the direct search
for SUSY particles has not been successful.
One can only set lower bounds of \order{100 \gev} on their
masses~\cite{pdg}.


\subsection{The MSSM Higgs sector}

An excellent review on this subject is given in \citere{awb2}.

\subsubsection{The Higgs boson sector at tree-level}
\label{sec:Higgstree}

Contrary to the Standard Model (SM), in the MSSM two Higgs doublets
are required.
The  Higgs potential~\cite{hhg}
\begin{align}
V &= m_{1}^2 |\cHe|^2 + m_{2}^2 |\cHz|^2 
      - m_{12}^2 (\epsilon_{ab} \cHe^a\cHz^b + \mbox{h.c.})  \non \\
  &  + \frac{1}{8}(g^2+g^{\prime 2}) \left[ |\cHe|^2 - |\cHz|^2 \right]^2
        + \frac{1}{2} g^2|\cHe^{\dag} \cHz|^2~,
\label{higgspot}
\end{align}
contains $m_1, m_2, m_{12}$ as soft SUSY breaking parameters;
$g, g'$ are the $SU(2)$ and $U(1)$ gauge couplings, and 
$\epsilon_{12} = -1$.

The doublet fields $H_1$ and $H_2$ are decomposed  in the following way:
\begin{align}
\cHe &= \VL \cHe^0 \\[0.5ex] \cHe^- \VR \; = \; \VL v_1 
        + \frac{1}{\sqrt2}(\phi_1^0 - i\chi_1^0) \\[0.5ex] -\phi_1^- \VR~,  
        \non \\
\cHz &= \VL \cHz^+ \\[0.5ex] \cHz^0 \VR \; = \; \VL \phi_2^+ \\[0.5ex] 
        v_2 + \frac{1}{\sqrt2}(\phi_2^0 + i\chi_2^0) \VR~.
\label{higgsfeldunrot}
\end{align}
$\cHe$ gives mass to the down-type fermions, while $\cHz$ gives masses
to the up-type fermions.
The potential (\ref{higgspot}) can be described with the help of two  
independent parameters (besides $g$ and $g'$): 
$\Tb = v_2/v_1$ and $M_A^2 = -m_{12}^2(\Tb+\CTb)$,
where $M_A$ is the mass of the $\cp$-odd Higgs boson~$A$.

Which values can be expected for $\tb$? One natural choice would be 
$\tb \approx 1$, i.e.\ both vevs are about the same. On the other hand, 
one can argue that $v_2$ is responsible for the top quark mass, while
$v_1$ gives rise to the bottom quark mass. Assuming that their mass
differences comes largely from the vevs, while their Yukawa couplings
could be about the same. The natural value for $\tb$ would then be 
$\tb \approx \mt/\mb$. Consequently, one can expect
\begin{align}
\label{tbrange}
1 \lsim \tb \lsim 50~.
\end{align}

The diagonalization of the bilinear part of the Higgs potential,
i.e.\ of the Higgs mass matrices, is performed via the orthogonal
transformations 
\begin{align}
\label{hHdiag}
\VL H^0 \\[0.5ex] h^0 \VR &= \ML \Ca & \Sa \\[0.5ex] -\Sa & \Ca \MR 
\VL \phi_1^0 \\[0.5ex] \phi_2^0~, \VR  \\
\label{AGdiag}
\VL G^0 \\[0.5ex] A^0 \VR &= \ML \Cb & \Sbe \\[0.5ex] -\Sbe & \Cb \MR 
\VL \chi_1^0 \\[0.5ex] \chi_2^0 \VR~,  \\
\label{Hpmdiag}
\VL G^{\pm} \\[0.5ex] H^{\pm} \VR &= \ML \Cb & \Sbe \\[0.5ex] -\Sbe & 
\Cb \MR \VL \phi_1^{\pm} \\[0.5ex] \phi_2^{\pm} \VR~.
\end{align}
The mixing angle $\al$ is determined through
\begin{align}
\al = {\rm arctan}\KKL 
  \frac{-(\MA^2 + \MZ^2) \Sbe \Cb}
       {\MZ^2 \CQb + \MA^2 \SQb - m^2_{h,{\rm tree}}} \KKR~, ~~
 -\frac{\pi}{2} < \al < 0
\label{alphaborn}
\end{align}
with $m_{h, {\rm tree}}$ defined below in \refeq{mhtree}.\\
One gets the following Higgs spectrum:
\begin{align}
\mbox{2 neutral bosons},\, {\cal CP} = +1 &: h, H \non \\
\mbox{1 neutral boson},\, {\cal CP} = -1  &: A \non \\
\mbox{2 charged bosons}                   &: H^+, H^- \non \\
\mbox{3 unphysical Goldstone bosons}      &: G, G^+, G^- .
\end{align}

At tree level the mass matrix of the neutral $\cp$-even Higgs bosons
is given in the $\Pe$-$\Pz$-basis 
in terms of $\MZ$, $\MA$, and $\Tb$ by
\begin{align}
M_{\rm Higgs}^{2, {\rm tree}} &= \ML \mpe^2 & \mpez^2 \\ 
                           \mpez^2 & \mpz^2 \MR \non\\
&= \ML \MA^2 \SQb + \MZ^2 \CQb & -(\MA^2 + \MZ^2) \Sbe \Cb \\
    -(\MA^2 + \MZ^2) \Sbe \Cb & \MA^2 \CQb + \MZ^2 \SQb \MR,
\label{higgsmassmatrixtree}
\end{align}
which by diagonalization according to \refeq{hHdiag} yields the
tree-level Higgs boson masses
\begin{align}
M_{\rm Higgs}^{2, {\rm tree}} 
   \stackrel{\al}{\longrightarrow}
   \ML m_{H,{\rm tree}}^2 & 0 \\ 0 &  m_{h,{\rm tree}}^2 \MR
\end{align}
with
\begin{align}
m_{H,h, {\rm tree}}^2 &= 
\edz \KKL \MA^2 + \MZ^2
         \pm \sqrt{(\MA^2 + \MZ^2)^2 - 4 \MZ^2 \MA^2 \CQZb} \KKR ~.
\label{mhtree}
\end{align}
From this formula the famous tree-level bound
\begin{align}
m_{h, {\rm tree}} \le \mbox{min}\{\MA, \MZ\} \cdot |\CZb| \le \MZ
\end{align}
can be obtained.
The charged Higgs boson mass is given by
\begin{align}
\label{rMSSM:mHp}
\mHp^2 = \MA^2 + \MW^2~.
\end{align}
The masses of the gauge bosons are given in analogy to the SM:
\begin{align}
M_W^2 = \frac{1}{2} g^2 (v_1^2+v_2^2) ;\qquad
M_Z^2 = \frac{1}{2}(g^2+g^{\prime 2})(v_1^2+v_2^2) ;\qquad M_\gamma=0.
\end{align}

\bigskip
The couplings of the Higgs bosons are modified from the corresponding SM
couplings already at the tree-level. Some examples are
\begin{align}
g_{hVV} &= \sin(\be - \al) \; g_{HVV}^{\rm SM}, \quad V = W^{\pm}, Z~, \\
g_{HVV} &= \cos(\be - \al) \; g_{HVV}^{\rm SM} ~,\\
g_{h b\bar b}, g_{h \tau^+\tau^-} &= - \frac{\sin\al}{\cos\be} \; 
                         g_{H b\bar b, H \tau^+\tau^-}^{\rm SM} ~, \\
g_{h t\bar t} &= \frac{\cos\al}{\sin\be} \; g_{H t\bar t}^{\rm SM} ~, \\
g_{A b\bar b}, g_{A \tau^+\tau^-} &= \ga_5\tb \; 
             g_{H b\bar b, H \tau^+\tau^-}^{\rm SM}~.
\end{align}
The following can be observed: the couplings of the $\cp$-even Higgs
boson to SM gauge bosons is always suppressed with respect to the SM
coupling. However, if $g_{hVV}^2$ is close to zero, $g_{HVV}^2$  is
close to $(g_{HVV}^{\rm SM})^2$ and vice versa, i.e.\ it is not possible
to decouple both of them from the SM gauge bosons. 
The coupling of the $h$ to down-type fermions can be suppressed 
{\em or enhanced} with respect to the SM value, depending on the size of
$\Sa/\Cb$. Especially for not too large values of $\MA$ and large $\tb$
one finds $|\Sa/\Cb| \gg 1$, leading to a strong enhancement of this
coupling. The same holds, in principle, for the coupling of the $h$ to
up-type fermions. However, for large parts of the MSSM parameter space
the additional factor is found to be $|\Ca/\Sbe| < 1$. For the $\cp$-odd
Higgs boson an additional factor $\tb$ is found. According to
\refeq{tbrange} this can lead to a strongly enhanced coupling of the
$A$~boson to bottom quarks or $\tau$~leptons, resulting in new search
strategies at the Tevatron and the LHC for the $\cp$-odd Higgs
boson, see \refse{sec:MSSMHiggsLHC}.

For $\MA \gsim 150 \gev$ the ``decoupling limit'' is reached. The
couplings of the light Higgs boson become SM-like, i.e.\ the additional
factors approach~1. The couplings of the heavy neutral Higgs bosons
become similar, $g_{Axx} \approx g_{Hxx}$, and the masses of the heavy
neutral and charged Higgs bosons fulfill $\MA \approx \MH \approx \MHp$. 
As a consequence, search strategies for the $A$~boson can also be
applied to the $H$~boson, and both are hard to disentangle at hadron
colliders.


\subsubsection{The scalar quark sector}
\label{sec:squark}

Since the most relevant squarks for the MSSM Higgs boson sector are
the $\Stop$~and $\Sbot$~particles, here we explicitly list 
their mass matrices in the basis of the gauge eigenstates 
$\StopL, \StopR$ and $\SbotL, \SbotR$:
\begin{align}
\label{stopmassmatrix}
{\cal M}^2_{\Stop} &=
  \ML \MstL^2 + \mt^2 + \CZb (\edz - \frac{2}{3} \sw^2) \MZ^2 &
      \mt \Xt \\
      \mt \Xt &
      \MstR^2 + \mt^2 + \frac{2}{3} \CZb \sw^2 \MZ^2 
  \MR, \\
& \non \\
\label{sbotmassmatrix}
{\cal M}^2_{\Sbot} &=
  \ML \MsbL^2 + \mb^2 + \CZb (-\edz + \frac{1}{3} \sw^2) \MZ^2 &
      \mb \Xb \\
      \mb \Xb &
      \MsbR^2 + \mb^2 - \frac{1}{3} \CZb \sw^2 \MZ^2 
  \MR.
\end{align}
$\MstL$, $\MstR$, $\MsbL$ and $\MsbR$ are the (diagonal) soft
SUSY-breaking parameters. We furthermore have
\begin{align}
\mt \Xt = \mt (\At - \mu \CTb) , \quad
\mb\, \Xb = \mb\, (\Ab - \mu \Tb) .
\label{eq:Xtb}
\end{align}
The soft SUSY-breaking parameters $\At$ and $\Ab$ denote the trilinear
Higgs--stop and Higgs--sbottom coupling, 
and $\mu$ is the Higgs mixing parameter.
$SU(2)$ gauge invariance requires the relation
\begin{align}
\MstL = \MsbL .
\end{align}
Diagonalizing ${\cal M}^2_{\Stop}$ and ${\cal M}^2_{\Sbot}$ with the
mixing angles $\tst$ and $\tsb$, respectively, yields the physical
$\Stop$~and $\Sbot$~masses: $\mste$, $\mstz$, $\msbe$ and $\msbz$.


\subsubsection{Higher-order corrections to Higgs boson masses}

A review about this subject can be found in \citere{habilSH}.
In the Feynman diagrammatic (FD) approach the higher-order corrected 
$\cp$-even Higgs boson masses in the rMSSM are derived by finding the
poles of the $(h,H)$-propagator 
matrix. The inverse of this matrix is given by
\begin{equation}
\left(\Delta_{\rm Higgs}\right)^{-1}
= - i \ML p^2 -  \mHtree^2 + \hSi_{HH}(p^2) &  \hSi_{hH}(p^2) \\
     \hSi_{hH}(p^2) & p^2 -  \mhtree^2 + \hSi_{hh}(p^2) \MR~.
\label{higgsmassmatrixnondiag}
\end{equation}
Determining the poles of the matrix $\Delta_{\rm Higgs}$ in
\refeq{higgsmassmatrixnondiag} is equivalent to solving
the equation
\begin{equation}
\left[p^2 - \mhtree^2 + \hSi_{hh}(p^2) \right]
\left[p^2 - \mHtree^2 + \hSi_{HH}(p^2) \right] -
\left[\hSi_{hH}(p^2)\right]^2 = 0\,.
\label{eq:proppole}
\end{equation}
The very leading one-loop correction to $\Mh^2$ is given by
\begin{align}
\De\Mh^2 &= \GF \mt^4 \log\KL\frac{\mste \mstz}{\mt^2}\KR~,
\label{DeltaMhmt4}
\end{align}
where $\GF$ denotes the Fermi constant. The \refeq{DeltaMhmt4} shows two
important aspects: First, the leading loop corrections go with $\mt^4$,
which is a ``very large number''. Consequently, the loop corrections can
strongly affect $\Mh$ and push the mass beyond the reach of
LEP~\cite{LEPHiggsSM,LEPHiggsMSSM}. Second, the scalar fermion masses
(in this case the scalar top masses) appear in the log entering the loop
corrections (acting as a ``cut-off'' where the new physics enter). In this
way the light Higgs boson mass depends on all other sectors via loop
corrections. This dependence is particularly pronounced for the scalar
top sector due to the large mass of the top quark.

The status of the available results for the self-energy contributions to
\refeq{higgsmassmatrixnondiag} can be summarized as follows. For the
one-loop part, the complete result within the MSSM is 
known~\cite{ERZ,mhiggsf1lA,mhiggsf1lB,mhiggsf1lC}. The by far dominant
one-loop contribution is the \order{\alt} term due to top and stop 
loops, see also \refeq{DeltaMhmt4}, 
($\alt \equiv h_t^2 / (4 \pi)$, $h_t$ being the superpotential top coupling).
Concerning the two-loop
effects, their computation is quite advanced and has now reached a
stage such that all the presumably dominant
contributions are known. They include the strong corrections, usually
indicated as \order{\alt\als}, and Yukawa corrections, \order{\alt^2},
to the dominant one-loop \order{\alt} term, as well as the strong
corrections to the bottom/sbottom one-loop \order{\alb} term ($\alb
\equiv h_b^2 / (4\pi)$), i.e.\ the \order{\alb\als} contribution. The
latter can be relevant for large values of $\Tb$. Presently, the
\order{\alt\als}~\cite{mhiggsEP1b,mhiggsletter,mhiggslong,mhiggsEP0,mhiggsEP1},
\order{\alt^2}~\cite{mhiggsEP1b,mhiggsEP3,mhiggsEP2} and the
\order{\alb\als}~\cite{mhiggsEP4,mhiggsFD2} contributions to the self-energies
are known for vanishing external momenta.  In the (s)bottom
corrections the all-order resummation of the $\Tb$-enhanced terms,
\order{\alb(\als\tb)^n}, is also performed \cite{deltamb1,deltamb}.
The \order{\alt\alb} and \order{\alb^2} corrections
were presented in~\citere{mhiggsEP4b}. A ``nearly full'' two-loop
effective potential calculation 
(including even the momentum dependence for the leading
pieces and the leading three-loop corrections) has been
published~\cite{mhiggsEP5}. 
Most recently another leading three-loop
calculation, valid for certain SUSY mass combinations, became
available~\cite{mhiggsFD3l}. 
Taking the available loop corrections into account, the upper limit of
$\Mh$ is shifted to~\cite{mhiggsAEC},
\begin{align}
\label{Mh135}
\Mh \le 135 \gev~
\end{align}
(as obtained with the code 
{\tt FeynHiggs}~\cite{feynhiggs,mhiggslong,mhiggsAEC,mhcMSSMlong}).
This limit takes into account the experimental uncertainty for the top
quark mass, see \refeq{mtexp}, as well as the intrinsic uncertainties
from unknown higher-order corrections~\cite{mhiggsAEC,PomssmRep}.

\medskip
The charged Higgs boson mass is obtained by solving the equation
\begin{align}
\label{rMSSM:mHpHO}
p^2 - \mHp^2 - \ser{H^-H^+}(p^2) = 0~.
\end{align}
The charged Higgs boson self-energy is known at the one-loop
level~\cite{chargedmhiggs,markusPhD}.


\subsection{MSSM Higgs boson searches at the LHC}
\label{sec:MSSMHiggsLHC}

The ``decoupling limit'' has been discussed for the tree-level couplings
and masses of the MSSM Higgs bosons in \refse{sec:Higgstree}.
This limit also persists taking into account radiative
corrections. The corresponding Higgs boson masses are shown in
\reffi{fig:decoupling} for $\tb = 5$ in the \mhmax~benchmark
scenario~\cite{benchmark2} obtained with {\tt FeynHiggs}.
For $\MA \gsim 150 \gev$ the lightest Higgs 
boson mass approaches its upper limit (depending on the SUSY
parameters), and the heavy Higgs boson masses are nearly degenerate.
Furthermore, also the light Higgs boson couplings including loop
corrections approach their SM-value for. Consequently, for $\MA \gsim 150
\gev$ the experimental 
searches for the lightest MSSM Higgs boson, see
\refse{sec:SMHiggsLHC}, can be performed very
similarly to the SM Higgs boson searches (with $\MHSM = \Mh$).

\begin{figure}[htb!]
\centerline{\includegraphics[width=.85\textwidth,height=8cm]{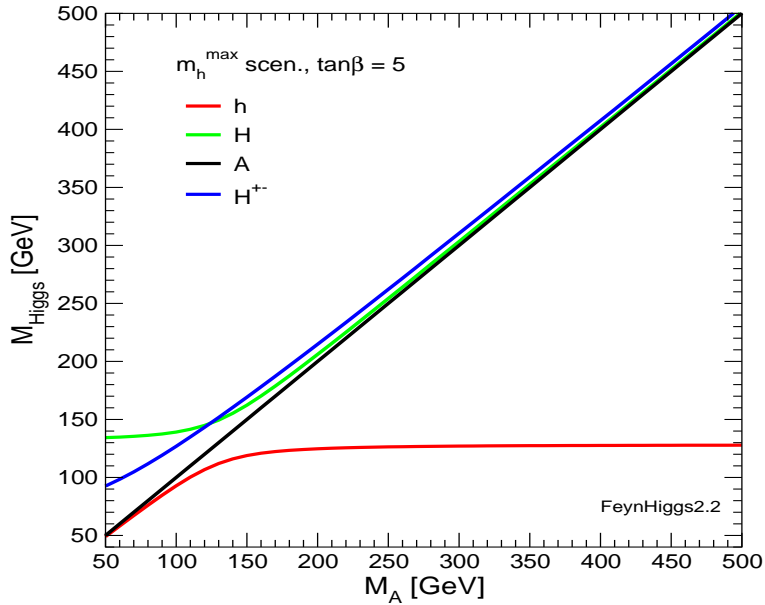}}
\caption{%
The MSSM Higgs boson masses including higher-order corrections are shown
as a function of $\MA$ for $\tb = 5$ in the \mhmax~benchmark
scenario~\cite{benchmark2} (obtained with 
{\tt FeynHiggs}~\cite{feynhiggs,mhiggslong,mhiggsAEC,mhcMSSMlong}).
}
\label{fig:decoupling}
\end{figure}

The various productions cross sections at the LHC are shown in
\reffi{fig:LHC_MSSM_XS} (for $\sqrt{s} = 14 \tev$). For low masses the
light Higgs cross sections are visible, and for $\MH \gsim 130 \gev$ the
heavy $\cp$-even Higgs cross section is displayed, while the cross
sections for the $\cp$-odd $A$~boson are given for the whole mass
range. As discussed in \refse{sec:Higgstree} the $g_{Abb}$ coupling is
enhanced by $\tb$ with respect to the corresponding SM
value. Consequently, the $b\bar b A$ cross section is the largest or
second largest cross section for all $\MA$, despite the 
relatively small value of $\tb = 5$. For larger $\tb$, see
\refeq{tbrange}, this cross section can become even more dominant.
Furthermore, the coupling of the
heavy $\cp$-even Higgs boson becomes very similar to the one of the
$A$~boson, and the two production cross sections, $b \bar b A$ and 
$b \bar b H$ are indistinguishable in the plot for $\MA > 200 \gev$.

\begin{figure}[htb!]
\begin{center}
\includegraphics[width=.85\textwidth,height=8cm]{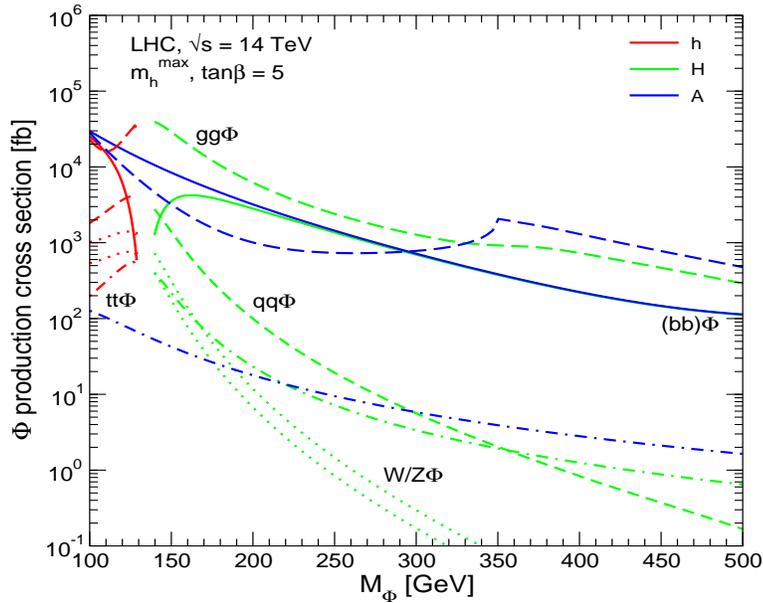}
\caption{%
Overview about the various neutral Higgs boson production cross sections
at the LHC shown as a function of $\MA$ for $\tb = 5$ in the \mhmax\
scenario (taken from \citere{sigmaH}, where the original references can
be found).
}
\label{fig:LHC_MSSM_XS}
\end{center}
\vspace{-1em}
\end{figure}

Following the above discussion, the main search channel for heavy Higgs
bosons at the LHC for 
$\MA \gsim 200 \gev$ is the production in association with bottom quarks
and the subsequent decay to tau leptons, 
$b \bar b \to b \bar b \; H/A \to b \bar b \; \tau^+\tau^-$.
For heavy supersymmetric particles, with masses far above
the Higgs boson mass scale, one has for the production and decay of the
$A$~boson~\cite{benchmark3} 
\begin{align}
\label{eq:bbA}
& \sigma(b\bar{b} A) \times {\rm BR}(A \to b \bar{b}) \simeq
\sigma(b\bar{b} H)_{\rm SM} \;
\frac{\tan^2\be}{\left(1 + \db \right)^2} \times
\frac{ 9}{
\left(1 + \db \right)^2 + 9} ~, \\
\label{eq:Atautau}
& \sigma(gg, b\bar{b} \to A) \times {\rm BR}(A \to \tau^+ \tau^-) \simeq
\sigma(gg, b\bar{b} \to H)_{\rm SM} \;
\frac{\tan^2\be}{
\left(1 + \db \right)^2 + 9} ~,
\end{align} 
where $\sigma(b\bar{b}H)_{\rm SM}$ and $\sigma(gg, b\bar{b} \to H)_{\rm SM}$ 
denote the values of the corresponding SM Higgs boson production cross
sections for $\MHSM = \MA$.
$\db$ is given by~\cite{deltamb1}
\BE
\db = \frac{2\als}{3\,\pi} \, \mgl \, \mu \, \tb \,
                    \times \, I(\msbe, \msbz, \mgl) +
      \frac{\alt}{4\,\pi} \, \At \, \mu \, \tb \,
                    \times \, I(\mste, \mstz, |\mu|) ~,
\label{def:dmb}
\end{equation}
where the function $I$ arises from the one-loop vertex diagrams and
scales as
$I(a, b, c) \sim 1/\mbox{max}(a^2, b^2, c^2)$.
Here $\mgl$ is the gluino mass, and $\mu$ is the Higgs mixing parameter.
As a consequence, the $b\bar{b}$ production rate depends sensitively on
$\db \propto \mu\,\tb$ because of the factor $1/(1 + \db)^2$, while this
leading 
dependence on $\db$ cancels out in the $\tau^+\tau^-$ production rate.
The formulas above apply, within a good approximation, also to the
heavy $\cp$-even Higgs boson in the large $\tb$ regime. 
Therefore, the production and decay
rates of $H$ are governed by similar formulas as the ones given
above, leading to an approximate enhancement by a factor 2 of the production
rates with respect to the ones that would be obtained in the case of the
single production of the $\cp$-odd Higgs boson as given in
\refeqs{eq:bbA}, (\ref{eq:Atautau}). 

Of particular interest is the ``LHC wedge'' region, i.e.\
the region in which only the light $\cp$-even MSSM Higgs boson, but non
of the heavy MSSM Higgs bosons can be
detected at the LHC at the 5$\,\si$ level. It appears for 
$\MA \gsim 200 \gev$ at intermediate $\tb$ and widens to larger $\tb$
values for larger $\MA$. Consequently, in the ``LHC wedge'' only a
SM-like light Higgs 
boson can be discovered at the LHC. This region is bounded from above by
the $5\,\si$ discovery contours for the heavy neutral MSSM Higgs bosons
as described above. These discovery contours depend sensitively 
on the Higgs mass parameter $\mu$. The dependence on
$\mu$ enters in two different ways, on the one hand via higher-order
corrections through $\db \propto \mu\,\tb$, 
and on the other hand
via the kinematics of Higgs decays into charginos and neutralinos, where
$\mu$ enters in their respective mass matrices~\cite{mssm}.

\begin{figure}[htb!]
\begin{center}
\includegraphics[width=.45\textwidth]{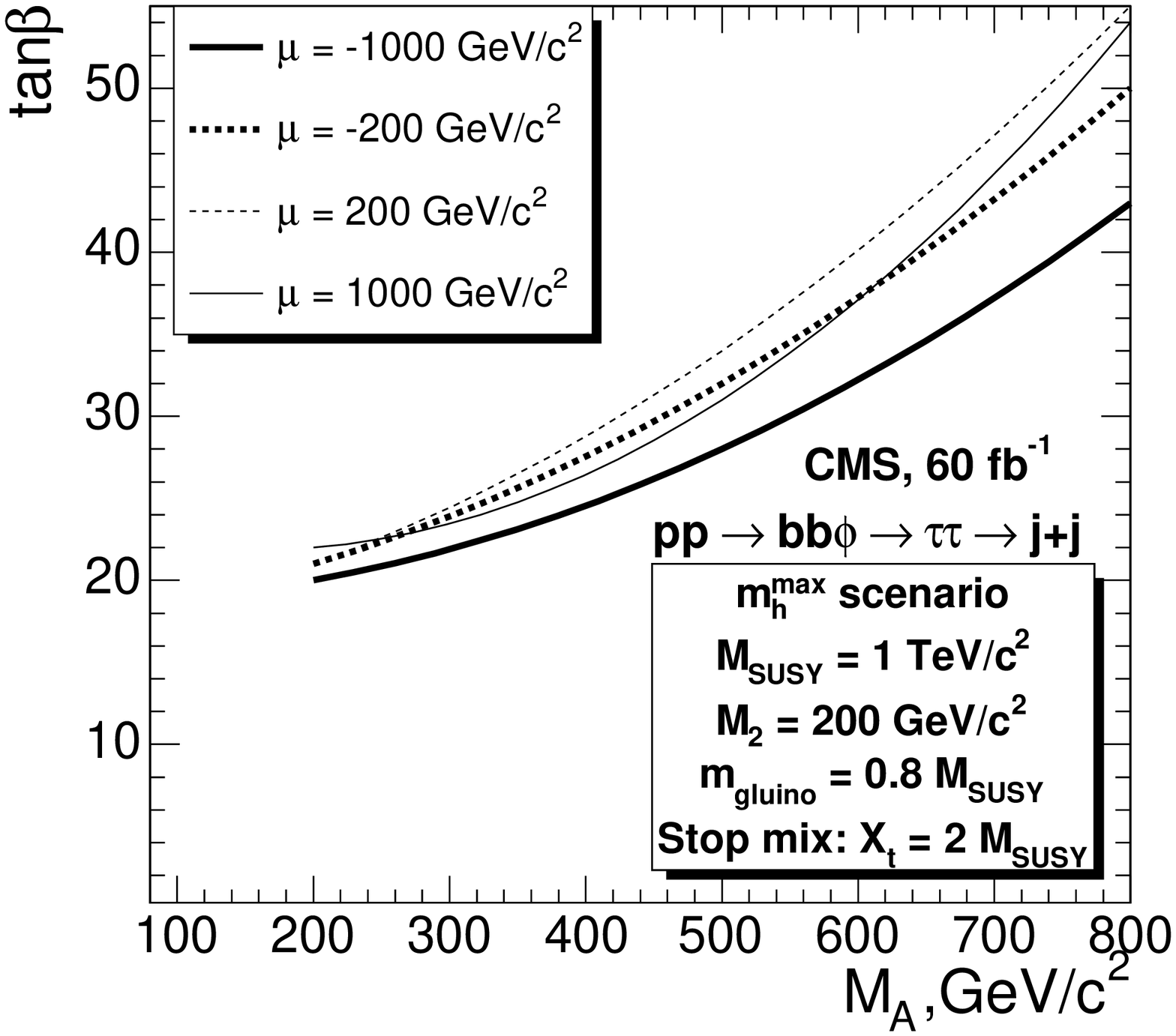}
\includegraphics[width=.45\textwidth]{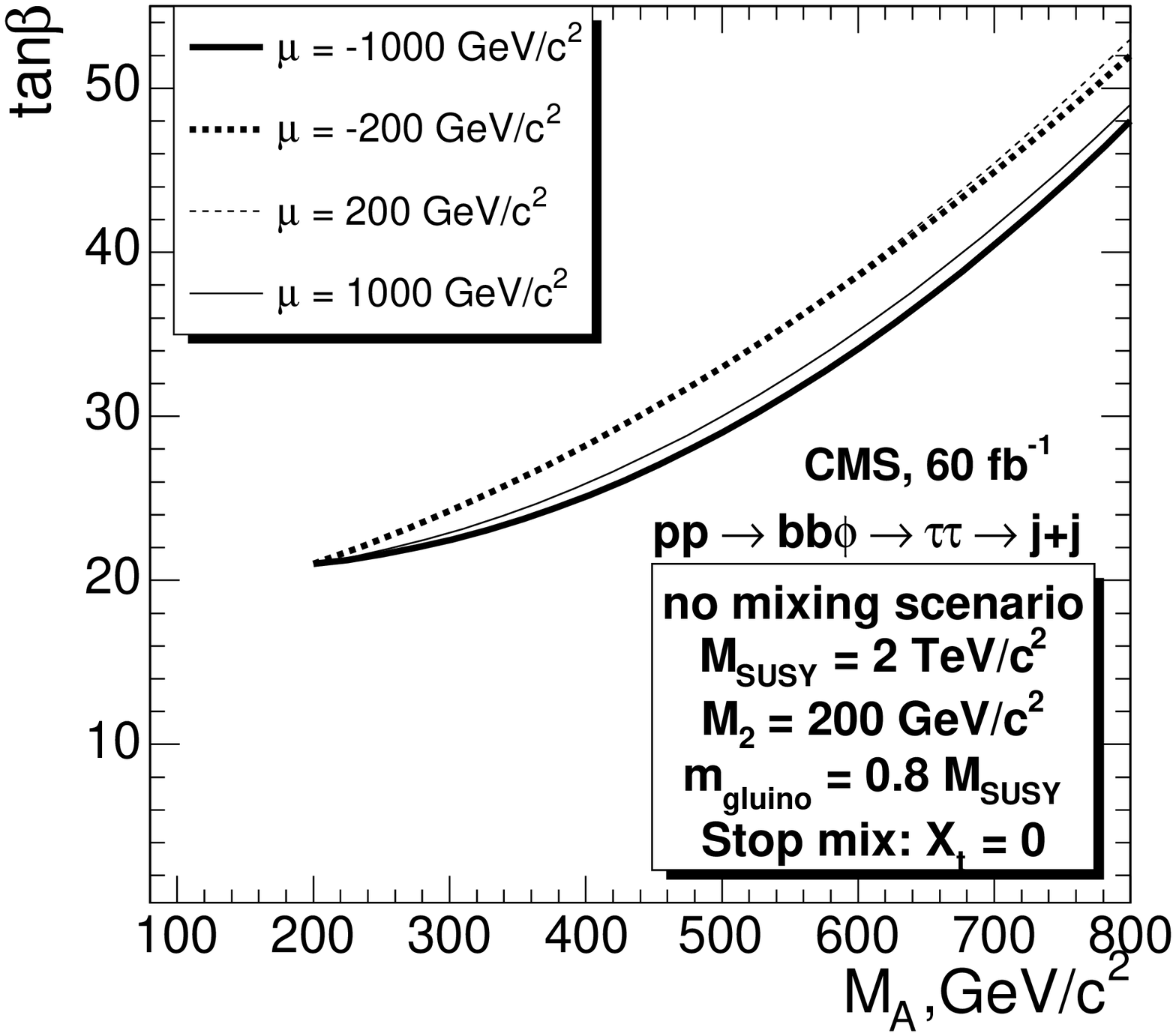}
\caption{%
The $5\,\si$ discovery regions (i.e.\ the upper bound of the ``LHC
wedge'' region) for the heavy neutral Higgs bosons in the channel 
$b \bar b \to b \bar b \; H/A , H/A \to \tau^+\tau^- \to \mbox{jets}$
(taken from \citere{cmsHiggs}).
}
\label{fig:LHCwedge}
\end{center}
\end{figure}

In \reffi{fig:LHCwedge} we show the $5\,\si$ discovery regions for the
heavy neutral MSSM Higgs bosons in the channel
$b \bar b \to b \bar b \; H/A , H/A \to \tau^+\tau^- \to \mbox{jets}$
\cite{cmsHiggs}. 
As explained above, these discovery contours correspond to the upper
bound of the ``LHC wedge''. A strong variation with the sign and the
size of $\mu$ can be observed and should be taken into account in
experimental and phenomenological analyses. 
The same higher-order corrections are relevant once a possible heavy
Higgs boson signal at the LHC will be interpreted in terms of the underlying
parameter space. From \refeq{def:dmb} it follows that an observed
production cross section can be correctly connected to $\mu$ and $\tb$
only if the scalar top and bottom masses, the gluino mass and the
trilinear Higgs-stop coupling are measured and taken properly into account.


\subsection{Electroweak precision observables}

Also within SUSY one can attempt to fit the unknown parameters to the
existing experimental data, in a similar fashion as it was discussed in
\refse{sec:ewpo}.
However, fits within the MSSM differs from the SM fit in various ways. First, 
the number of free parameters is substantially larger in the MSSM, even
restricting to GUT based models as discussed below.
On the other hand, more observables can be taken into account, providing
extra constraints on the fit. Within the MSSM the additional observables 
included are the anomalous magnetic moment of the muon $(g-2)_\mu$,
$B$-physics observables such as $\br(b \to s \ga)$ or 
$\br(B_s \to \mu\mu)$, and the relic density of cold dark matter (CDM),
which can be provided by the lightest SUSY particle, the neutralino. 
These additional constraints would either have a minor impact on
the best-fit regions or cannot be accommodated in the SM.
Finally, as discussed in the previous subsections, whereas the light
Higgs boson mass is a free parameter 
in the SM, it is a function of the other parameters in the MSSM.
In this way, for example, the masses of the scalar tops and
bottoms enter not only directly into the prediction of the various
observables, but also indirectly via their impact on $\Mh$.

Within the MSSM the dominant SUSY correction to electroweak precision
observables arises from 
the scalar top and bottom contribution to the $\rho$~parameter,
see~\refeq{delrho}. The leading diagrams are shown in 
\reffi{fig:fdgb1l}. 

\begin{figure}[htb!]
\begin{center}
\mbox{
\psfig{figure=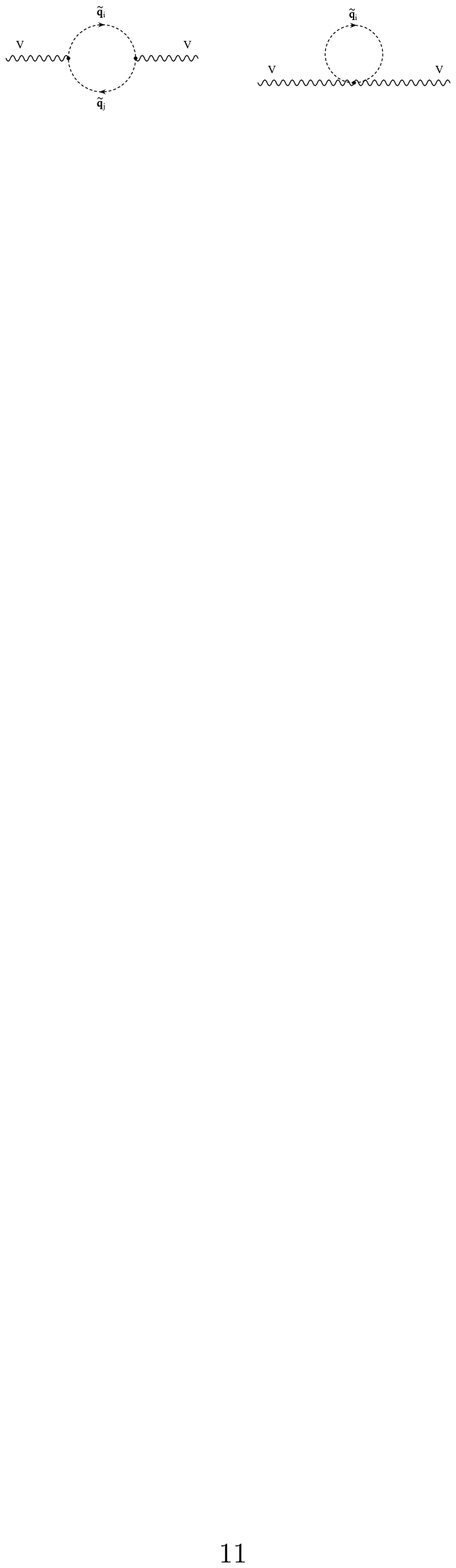,width=11cm,
       bbllx=210pt,bblly=680pt,bburx=397pt,bbury=720pt}}
\end{center}
\caption[]{Feynman diagrams for the contribution of scalar quark loops 
to the gauge boson self-energies at one-loop order, $V = W, Z$, 
$\sq = \Stop, \Sbot$.}
\label{fig:fdgb1l}
\end{figure}

Generically one finds $\De\rho^{\SU} > 0$, leading,
for instance, to an upward shift in the prediction of $\MW$ with respect
to the SM prediction. 
The experimental result and the theory prediction of the SM and the MSSM
for $\MW$ are compared in \reffi{fig:MWMTtoday} (updated from
\citere{Heinemeyer:2006px}). 
The predictions within the two models 
give rise to two bands in the $\mt$--$\MW$ plane with only a relatively small
overlap sliver (indicated by a dark-shaded (blue) area in
\reffi{fig:MWMTtoday}).  
The allowed parameter region in the SM (the medium-shaded (red)
and dark-shaded (blue) bands, corresponding to the SM prediction in
\reffi{fig:MWMTSM}) arises from varying the only free parameter  
of the model, the mass of the SM Higgs boson, from $\MHSM = 114\gev$,
the LEP exclusion bound~\cite{LEPHiggsSM}
(upper edge of the dark-shaded (blue) area), to $400 \gev$ (lower edge
of the medium-shaded (red) area).
The light shaded (green) and the dark-shaded (blue) areas indicate 
allowed regions for the unconstrained MSSM, obtained from scattering the
relevant parameters independently~\cite{Heinemeyer:2006px}. 
The decoupling limit with SUSY masses of \order{2 \tev}
yields the lower edge of the dark-shaded (blue) area. Thus, the overlap 
region between
the predictions of the two models corresponds in the SM to the region
where the Higgs boson is light, i.e.\ in the MSSM allowed region 
($\Mh \lsim 135 \gev$, see \refeq{Mh135}). In the MSSM it
corresponds to the case where all 
superpartners are heavy, i.e.\ the decoupling region of the MSSM.
The current 68~and~95\%~C.L.\ experimental results 
for $\mt$, \refeq{mtexp}, and $\MW$, \refeq{MWexp}, are also indicated
in the plot. As can be seen from 
\reffi{fig:MWMTtoday}, the current experimental 68\%~C.L.\ region for 
$\mt$ and $\MW$ exhibits a slight preference of the MSSM over the SM.
This example indicates that the experimental measurement of $\MW$
in combination with $\mt$ 
prefers, within the MSSM, not too heavy SUSY mass scales.

\begin{figure}[htb!]
\begin{center}
\includegraphics[width=.65\textwidth]{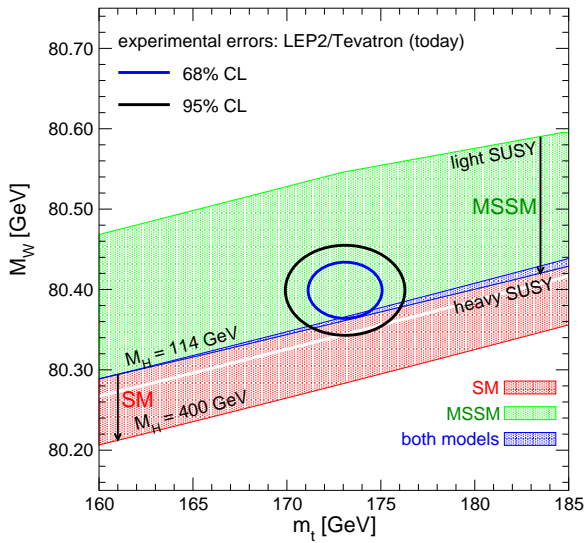}
\begin{picture}(0,0)
\CBox(-130,024)(-13,030){White}{White}
\end{picture}
\vspace{-1em}
\caption{%
Prediction for $\MW$ in the MSSM and the SM (see text) as a function of
$\mt$ in comparison with the present experimental results for $\MW$ and
$\mt$ (updated from \citere{Heinemeyer:2006px}, see \citere{PomssmRep}
for details).
}
\label{fig:MWMTtoday}
\end{center}
\end{figure}

\bigskip
As mentioned above, in order to restrict the number of free parameters
in the MSSM one can resort to GUT based models. Most fits have been
performed in the Constrained MSSM (CMSSM), in
which the input scalar masses $m_0$, gaugino masses $m_{1/2}$ and
soft trilinear parameters $A_0$ are each universal at the GUT scale,
$M_{\rm GUT} \approx 2 \times 10^{16} \gev$, 
and in the Non-universal Higgs mass model (NUHM1), in which a common
SUSY-breaking contribution to the Higgs masses is allowed to be
non-universal. 

Here we follow the results obtained in
\citeres{Master1,Master2,Master3}, where an overview about different
fitting techniques and extensive list of references can be found in
\citere{Master3}. The computer code used for the fits shown below is the
{\tt MasterCode} \cite{Master1,Master2,Master3,MasterWWW},
which includes the following theoretical codes. For the RGE running of
the soft SUSY-breaking parameters, it uses
{\tt SoftSUSY}~\cite{Allanach:2001kg}, which is combined consistently
with the codes used for the various low-energy observables:
{\tt FeynHiggs}~\cite{feynhiggs,mhiggslong,mhiggsAEC,mhcMSSMlong}  
is used for the evaluation of the Higgs masses and  
$a_\mu^{\rm SUSY}$  (see also
\cite{Moroi:1995yh,Degrassi:1998es,Heinemeyer:2003dq,Heinemeyer:2004yq}),
for the other electroweak precision data we have included 
a code based on~\cite{Heinemeyer:2006px,Heinemeyer:2007bw},
{\tt SuFla}~\cite{Isidori:2006pk,Isidori:2007jw} and 
{\tt SuperIso}~\cite{Mahmoudi:2008tp,Eriksson:2008cx}
are used for flavor-related observables, 
and for dark-matter-related observables
{\tt MicrOMEGAs}~\cite{Belanger:2006is} and
{\tt DarkSUSY}~\cite{Gondolo:2005we} are used.
In the combination of the various codes,
{\tt MasterCode} makes extensive use of the SUSY
Les Houches Accord~\cite{Skands:2003cj,Allanach:2008qq}.

The global $\chi^2$ likelihood function, which combines all
theoretical predictions with experimental constraints, is now given as
\begin{align}
\chi^2 &= \sum^N_i \frac{(C_i - P_i)^2}{\sigma(C_i)^2 + \sigma(P_i)^2}
+ \sum^M_i \frac{(f^{\rm obs}_{{\rm SM}_i}
              - f^{\rm fit}_{{\rm SM}_i})^2}{\sigma(f_{{\rm SM}_i})^2}
\nonumber \\[.2em]
&+ \chi^2(\br(B_s \to \mu\mu))
+ \chi^2(\mbox{SUSY search limits})
\label{eqn:chi2}
\end{align} 
Here $N$ is the number of observables studied, $C_i$ represents an
experimentally measured value (constraint) and each $P_i$ defines a
prediction for the corresponding constraint that depends on the
supersymmetric parameters.
The experimental uncertainty, $\sigma(C_i)$, of each measurement is
taken to be both statistically and systematically independent of the
corresponding theoretical uncertainty, $\sigma(P_i)$, in its
prediction (all the details can be found in \citere{Master3}).
$\chi^2(\br(B_s \to \mu\mu))$ denotes the $\chi^2$
contributions from the one measurement for which only a one-sided
bound are available so far.
Furthermore included are the lower limits from the direct searches
for SUSY particles at LEP~\cite{LEPSUSY} as one-sided limits, denoted by 
``$\chi^2(\mbox{SUSY search limits})$'' in \refeq{eqn:chi2}.
Furthermore, the three SM parameters
$f_{\rm SM} = \{\Delta\alpha_{\rm had}, \mt, \MZ \}$ are included as fit
parameters and allowed to vary with their current experimental
resolutions $\sigma(f_{\rm SM})$. 

The results for the fits of $\Mh$ in the CMSSM and the NUHM1 are shown
in \reffi{fig:redband} in the left and right plot, respectively.
Also shown in
\reffi{fig:redband} are the LEP exclusion on a SM Higgs
(yellow shading)
and the ranges that are theoretically inaccessible in the
supersymmetric models studied (beige shading). 
The LEP exclusion is directly applicable to the CMSSM,
but cannot that strictly be applied in the NUHM1, see \citere{Master3}
for details.

\begin{figure}[htb!]
\vspace{-1em}
\includegraphics[width=.49\textwidth]{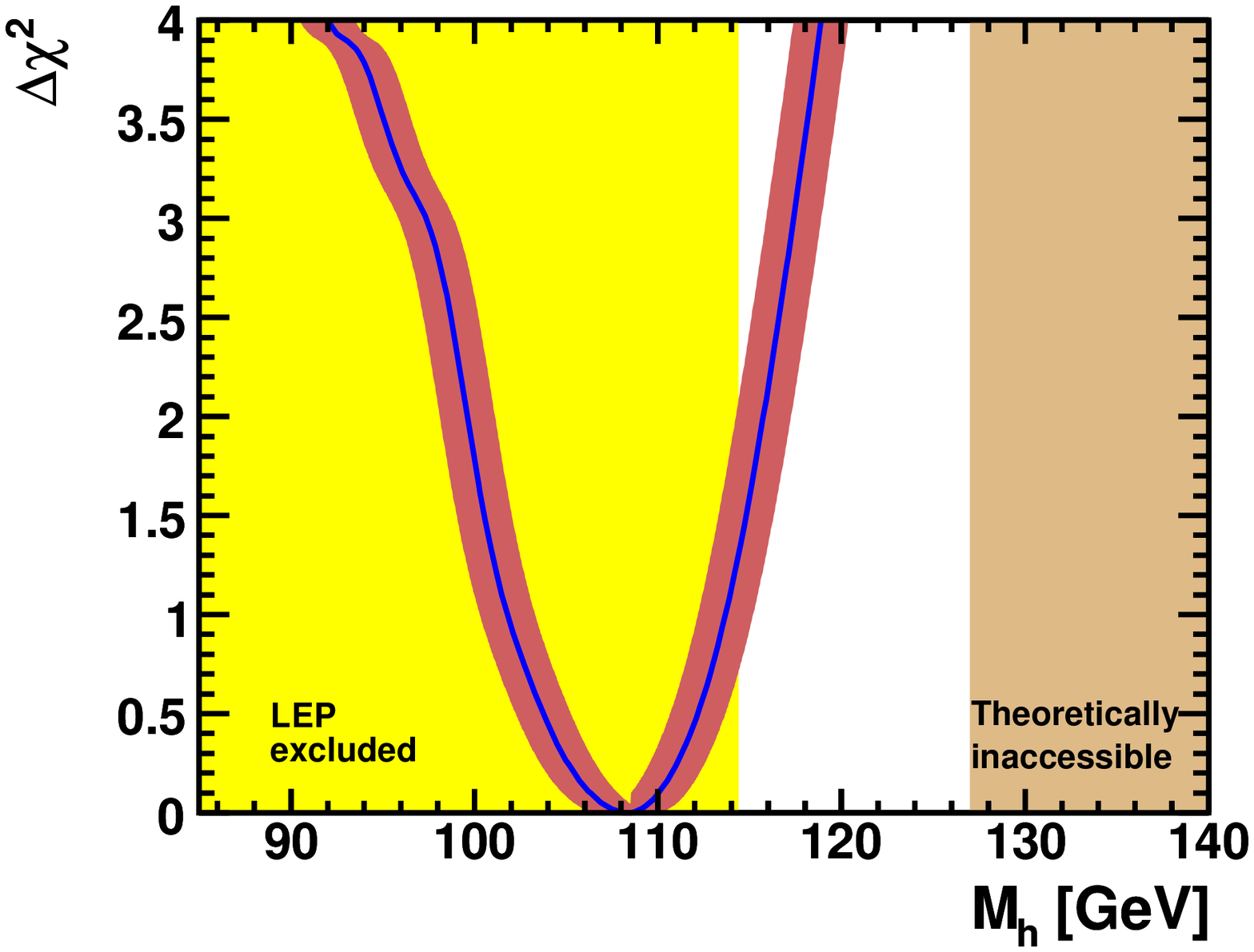}~~
\includegraphics[width=.49\textwidth]{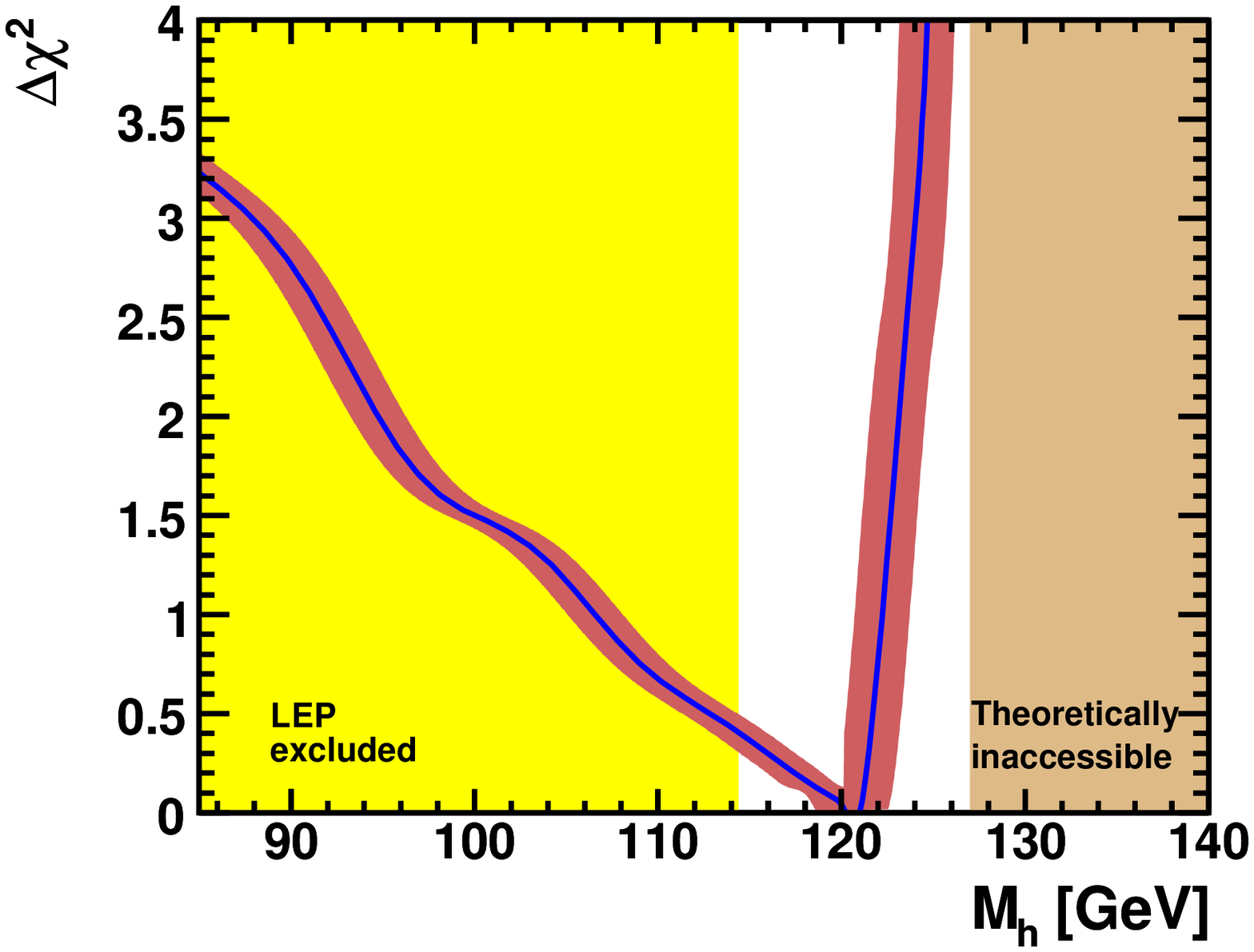}
\caption{%
The $\chi^2$ functions for $\Mh$ in the CMSSM (left) and
  the NUHM1 (right)~\cite{Master3}, 
including the theoretical uncertainties (red bands). Also shown is the mass
range excluded for a SM-like Higgs boson (yellow shading),
and the ranges theoretically inaccessible in the supersymmetric models
studied.}
\label{fig:redband}
\end{figure}

In the case of the CMSSM, we see in the left panel of
\reffi{fig:redband} that the minimum of the $\chi^2$
function occurs below the LEP exclusion limit. 
The fit result is still
compatible at the 95\% C.L.\ with the search limit, 
similarly to the SM case. 
In the case of the NUHM1, shown in the right panel of
Fig.~\ref{fig:redband}, we see that the minimum of the $\chi^2$
function occurs {\it above} the LEP lower limit on the mass of a SM 
Higgs. Thus, 
within the NUHM1 the combination of all other experimental
constraints {\em naturally} evades the LEP Higgs constraints, and no
tension between $\Mh$ and the experimental bounds exists.








\section*{Acknowledgments} 
I thank the organizers for their hospitality and for creating a very
stimulating environment, as well as, together with the
participants, for an exceptionally nice school dinner/farewell party.

\begin{susspbibliography}{99} 


\bibitem{higgs-mechanism}
        P.~W.~Higgs,
        Phys.\ Lett.\  {\bf 12} (1964) 132;
        Phys.\ Rev.\ Lett.\  {\bf 13} (1964) 508;
        Phys.\ Rev.\  {\bf 145} (1966) 1156;\\
        F.~Englert and R.~Brout,
        Phys.\ Rev.\ Lett.\  {\bf 13} (1964) 321;\\
        G.~S.~Guralnik, C.~R.~Hagen and T.~W.~B.~Kibble,
        Phys.\ Rev.\ Lett.\  {\bf 13} (1964) 585.

\bibitem{sm} S.L.~Glashow, 
             {\em Nucl.\ Phys.} {\bf B 22} (1961) 579; \\
             S. Weinberg, 
             {\em Phys. Rev. Lett.} {\bf 19} (1967) 19; \\ 
             A. Salam, in: {\em Proceedings of the 8th Nobel 
             Symposium}, Editor N. Svartholm, Stockholm, 1968.

\bibitem{mssm} H.~Nilles, 
               {\em Phys.\ Rept.} {\bf 110} (1984) 1; \\ 
               H.~Haber and G.~Kane, 
               {\em Phys.\ Rept.} {\bf 117} (1985) 75; \\  
               R.~Barbieri, 
               {\em Riv.\ Nuovo Cim.} {\bf 11} (1988) 1. 

\bibitem{thdm} S.~Weinberg,
        Phys.\ Rev.\ Lett.\  {\bf 37} (1976) 657;\\
        J.~Gunion, H.~Haber, G.~Kane and S.~Dawson,
        {\em The Higgs Hunter's Guide}
        (Perseus Publishing, Cambridge, MA, 1990),
        and references therein.

\bibitem{NMSSM-etc}
        P.~Fayet, 
        {\em Nucl. Phys.} {\bf B 90} (1975) 104;
        {\em Phys. Lett.} {\bf B 64} (1976) 159;
        {\em Phys. Lett.} {\bf B 69} (1977) 489;
        {\em Phys. Lett.} {\bf B 84} (1979) 416;\\
        H.P.~Nilles, M.~Srednicki and D.~Wyler,
        {\em Phys. Lett.} {\bf B 120} (1983) 346;\\
        J.M.~Frere, D.R.~Jones and S.~Raby,
        {\em Nucl. Phys.} {\bf B 222} (1983) 11;\\
        J.P.~Derendinger and C.A.~Savoy, 
        {\em Nucl. Phys.} {\bf B 237} (1984) 307;\\
        J.~Ellis, J.~Gunion, H.~Haber, L.~Roszkowski and F.~Zwirner,
        {\em Phys. Rev.} {\bf D 39}  (1989) 844;\\
        M.~Drees, 
        {\em Int. J. Mod. Phys.} {\bf A 4}  (1989) 3635. 

\bibitem{lhm} N.~Arkani-Hamed, A.~Cohen and H.~Georgi,
              {\em Phys.\ Lett.} {\bf B 513} (2001) 232
              [arXiv:hep-ph/0105239];\\
              N.~Arkani-Hamed, A.~Cohen, T.~Gregoire and J.~Wacker,
              {\em JHEP} {\bf 0208} (2002) 020
              [arXiv:hep-ph/0202089].

\bibitem{edm} N.~Arkani-Hamed, S.~Dimopoulos and G.~Dvali,
              {\em Phys.\ Lett.} {\bf B 429} (1998) 263
              [arXiv:hep-ph/9803315];
              {\em Phys.\ Lett.} {\bf B 436} (1998) 257
              [arXiv:hep-ph/9804398];\\
              I.~Antoniadis,
              {\em Phys.\ Lett.} {\bf B 246} (1990) 377;\\
              J.~Lykken,
              {\em Phys.\ Rev.} {\bf D 54} (1996) 3693
              [arXiv:hep-th/9603133];\\
              L.~Randall and R.~Sundrum,
              {\em Phys.\ Rev.\ Lett.} {\bf 83} (1999) 3370
              [arXiv:hep-ph/9905221]. 

\bibitem{lhcilc} G.~Weiglein et al.\ [LHC/ILC Study Group],
                 {\em Phys.\ Rept.} {\bf 426} (2006) 47
                 [arXiv:hep-ph/0410364].

\bibitem{lhc2fc} A.~De Roeck et al.,
                 arXiv:0909.3240 [hep-ph].

\bibitem{RGEla1} N.~Cabibbo, L.~Maiani, G.~Parisi and R.~Petronzio,
                 {\em Nucl.\ Phys.} {\bf B 158} (1979) 295;\\
                 R.~Flores and M.~Sher,
                 {\em Phys.\ Rev.} {\bf D 27} (1983) 1679;\\
                 M.~Lindner,
                 {\em Z.\ Phys.} {\bf C 31} (1986) 295;\\
                 M.~Sher,
                 {\em Phys.\ Rept.} {\bf 179} (1989) 273;\\
                 J.~Casas, J.~Espinosa and M.~Quiros,
                 {\em Phys.\ Lett.} {\bf 342} (1995) 171.
                 [arXiv:hep-ph/9409458].

\bibitem{RGEla2} G.~Altarelli and G.~Isidori,
                 {\em Phys.\ Lett.} {\bf B 337} (1994) 141;
                 J.~Espinosa and M.~Quiros,
                 {\em Phys.\ Lett.} {\bf 353} (1995) 257
                 [arXiv:hep-ph/9504241].

\bibitem{RGEla3} T.~Hambye and K.~Riesselmann,
                 {\em Phys.\ Rev.} {\bf D 55} (1997) 7255
                 [arXiv:hep-ph/9610272].

\bibitem{sigmaH} T.~Hahn, S.~Heinemeyer, F.~Maltoni, G.~Weiglein and 
                 S.~Willenbrock,
                 arXiv:hep-ph/0607308.

\bibitem{atlas}
  G.~Aad et al.\ [The ATLAS Collaboration],
  arXiv:0901.0512.

\bibitem{cms}
  G.~Bayatian et al.\ [CMS Collaboration],
  {\em J.\ Phys.} {\bf G 34} (2007) 995.

\bibitem{LEPEWWG} LEP Electroweak Working Group,\\
                  see: {\tt lepewwg.web.cern.ch/LEPEWWG/Welcome.html} .

\bibitem{TEVEWWG} Tevatron Electroweak Working Group,
                  see: {\tt tevewwg.fnal.gov} .

\bibitem{lepewwg} The ALEPH, DELPHI, L3, OPAL, SLD Collaborations,
                  the LEP Electroweak Working Group,
                  the SLD Electroweak and Heavy Flavour Groups,
                  {\em Phys.\ Rept.} {\bf 427} (2006) 257
                  [arXiv:hep-ex/0509008];\\
                  {}[The ALEPH, DELPHI, L3 and OPAL Collaborations, the LEP
                  Electroweak Working Group], 
                  arXiv:hep-ex/0612034.

\bibitem{sirlin} A.~Sirlin, 
                 {\em Phys. Rev.} {\bf D 22} (1980) 971;
                 W.~Marciano and A.~Sirlin, 
                 {\em Phys. Rev.} {\bf D 22} (1980) 2695.

\bibitem{rho} M.~Veltman, 
              {\em Nucl. Phys.} {\bf B 123} (1977) 89. 

\bibitem{LEPHiggsSM} LEP Higgs working group,
                     {\em Phys. Lett.} {\bf B 565} (2003) 61
                     [arXiv:hep-ex/0306033].

\bibitem{MW80399} ALEPH Collaboration, CDF Collaboration, D0
  Collaboration, DELPHI Collaboration, L3 Collaboration, OPAL
  Collaboration, SLD Collaboration, LEP Electroweak Working Group,
  Tevatron Electroweak Working Group, SLD electroweak heavy flavour
  groups, 
                     arXiv:0911.2604 [hep-ex].

\bibitem{mt1731} Tevatron Electroweak Working Group and CDF Collaboration
                 and D0 Collaboration,
                 arXiv:0903.2503 [hep-ex].

\bibitem{gruenewald07} M.~Gr\"unewald, 
                       arXiv:0709.3744 [hep-ex];
                       arXiv:0710.2838 [hep-ex].

\bibitem{gruenewaldpriv} M.~Gr\"unewald,
                         {\em priv. communication}.

\bibitem{topaz0} G.~Montagna, O.~Nicrosini, F.~Piccinini and G.~Passarino,
                 {\em Comput.\ Phys.\ Commun.} {\bf 117} (1999) 278
                 [arXiv:hep-ph/9804211].

\bibitem{zfitter} D.~Bardin et al., 
                  {\em Comput.\ Phys.\ Commun.} {\bf 133} (2001) 229
                  [arXiv:hep-ph/9908433];
                  A.~Arbuzov et al.,
                  {\em Comput.\ Phys.\ Commun.} {\bf 174} (2006) 728
                  [arXiv:hep-ph/0507146].

\bibitem{TevHiggsSM160170} [CDF Collaboration and D0 Collaboration],
                     arXiv:0903.4001 [hep-ex].

\bibitem{GFitter} H.~Flacher, M.~Goebel, J.~Haller, A.~Hocker, K.~Moenig
                  and J.~Stelzer, 
                  {\em Eur.\ Phys.\ J.} {\bf C 60} (2009) 543
                  [arXiv:0811.0009 [hep-ph]];\\
                  see: {\tt cern.ch/gfitter}~.

\bibitem{TevHiggsSM} [CDF Collaboration and D0 Collaboration],
                     arXiv:0911.3930 [hep-ex].

\bibitem{blueband} U.~Baur, R.~Clare, J.~Erler, S.~Heinemeyer,
                   D.~Wackeroth, G.~Weiglein and D.~Wood,
                   arXiv:hep-ph/0111314.

\bibitem{gigaz} J.~Erler, S.~Heinemeyer, W.~Hollik, G.~Weiglein 
                and P.~Zerwas,
                {\em Phys. Lett.} {\bf B 486} (2000) 125
                [arXiv:hep-ph/0005024];\\
                J.~Erler and S.~Heinemeyer,
                arXiv:hep-ph/0102083.

\bibitem{moenig} R.~Hawkings and K.~M\"onig, 
                 {\em EPJdirect} {\bf C8} (1999) 1
                 [arXiv:hep-ex/9910022].

\bibitem{mwgigaz} G.~Wilson, LC-PHSM-2001-009, see:
                  {\tt www.desy.de/$\sim$lcnotes/notes.html}.

\bibitem{Snowmass05Higgs} S.~Heinemeyer et al.,
                          arXiv:hep-ph/0511332.

\bibitem{glawei} S.~Glashow and S.~Weinberg,
                 {\em Phys. Rev.} {\bf D 15} (1977) 1958.

\bibitem{pdg} C.~Amsler et al.\ [Particle Data Group],
  {\em Phys.\ Lett.} {\bf B 667} (2008) 1.

\bibitem{awb2} A.~Djouadi,
               {\em Phys.\ Rept.} {\bf 459} (2008) 1
               [arXiv:hep-ph/0503173].

\bibitem{hhg} J.~Gunion, H.~Haber, G.~Kane and S.~Dawson,
              {\em The Higgs Hunter's Guide}, Addison-Wesley, 1990.

\bibitem{habilSH} S.~Heinemeyer, 
                  {\em Int. J. Mod. Phys.} {\bf A 21} (2006) 2659
                  [arXiv:hep-ph/0407244].

\bibitem{LEPHiggsMSSM} ~LEP Higgs working group,
                       {\em Eur.\ Phys.\ J.} {\bf C 47} (2006) 547
                       [arXiv:hep-ex/0602042].

\bibitem{ERZ} J.~Ellis, G.~Ridolfi and F.~Zwirner,
              {\em Phys.\ Lett.} {\bf B 257} (1991) 83;\\
              Y.~Okada, M.~Yamaguchi and T.~Yanagida,
              {\em Prog.\ Theor.\ Phys. } {\bf 85} (1991) 1;\\
              H.~Haber and R.~Hempfling,
              {\em Phys.\ Rev.\ Lett.}  {\bf 66} (1991) 1815.

\bibitem{mhiggsf1lA} A.~Brignole,
                     {\em Phys. Lett.}\ {\bf B 281} (1992) 284.

\bibitem{mhiggsf1lB} P.~Chankowski, S.~Pokorski and J.~Rosiek,
                     {\em Phys. Lett.} {\bf B 286} (1992) 307;
                     {\em Nucl. Phys.} {\bf B 423} (1994) 437,
                     hep-ph/9303309.

\bibitem{mhiggsf1lC} A.~Dabelstein,
                     {\em Nucl. Phys.} {\bf B 456} (1995) 25,
                     hep-ph/9503443;
                     {\em Z. Phys.} {\bf C 67} (1995) 495,
                     hep-ph/9409375.

\bibitem{mhiggsEP1b} R.~Hempfling and A.~Hoang, 
                     {\em Phys. Lett.} {\bf B 331} (1994) 99, 
                     hep-ph/9401219.

\bibitem{mhiggsletter} S.~Heinemeyer, W.~Hollik and G.~Weiglein, 
                       {\em Phys. Rev.} {\bf D 58} (1998) 091701, 
                       hep-ph/9803277; 
                       {\em Phys. Lett.} {\bf B 440} (1998) 296, 
                       hep-ph/9807423.

\bibitem{mhiggslong} S.~Heinemeyer, W.~Hollik and G.~Weiglein, 
                     {\em Eur. Phys. Jour.} {\bf C 9} (1999) 343, 
                     hep-ph/9812472.

\bibitem{mhiggsEP0}  R.~Zhang, {\em Phys.\ Lett. } {\bf B 447} (1999) 89, 
                     hep-ph/9808299;\\
                     J.~Espinosa and R.~Zhang, {\em JHEP} {\bf 0003} 
                     (2000) 026, hep-ph/9912236.

\bibitem{mhiggsEP1} G.~Degrassi, P.~Slavich and F.~Zwirner,
                    {\em Nucl. Phys.} {\bf B 611} (2001) 403,
                    hep-ph/0105096.

\bibitem{mhiggsEP3} J.~Espinosa and R.~Zhang,
                    {\em Nucl. Phys.} {\bf B 586} (2000) 3,
                    hep-ph/0003246. 

\bibitem{mhiggsEP2} A.~Brignole, G.~Degrassi, P.~Slavich and F.~Zwirner,
                    {\em Nucl. Phys.} {\bf B 631} (2002) 195,
                    hep-ph/0112177.

\bibitem{mhiggsEP4} A.~Brignole, G.~Degrassi, P.~Slavich and F.~Zwirner,
                    {\em Nucl. Phys.} {\bf B 643} (2002) 79,
                    hep-ph/0206101.

\bibitem{mhiggsFD2} S.~Heinemeyer, W.~Hollik, H.~Rzehak and G.~Weiglein,
                    {\em Eur. Phys. J.} {\bf C 39} (2005) 465
                    [arXiv:hep-ph/0411114].

\bibitem{deltamb1} T.~Banks, 
                   {\em Nucl.\ Phys.} {\bf B 303} (1988) 172;\\
                   L.~Hall, R.~Rattazzi and U.~Sarid,
                   {\em Phys.\ Rev.} {\bf D 50} (1994) 7048, 
                   hep-ph/9306309;\\
                   R.~Hempfling, 
                   {\em Phys.\ Rev.} {\bf D 49} (1994) 6168;\\
                   M.~Carena, M.~Olechowski, S.~Pokorski and C.~Wagner,
                   {\em Nucl.\ Phys.}\ {\bf B 426} (1994) 269, 
                   hep-ph/9402253.

\bibitem{deltamb} M.~Carena, D.~Garcia, U.~Nierste and C.~Wagner,
                  {\em Nucl. Phys.} {\bf B 577} (2000) 577,
                  hep-ph/9912516.

\bibitem{mhiggsEP4b} G.~Degrassi, A.~Dedes and P.~Slavich,
                    {\em Nucl. Phys.} {\bf B 672} (2003) 144, 
                    hep-ph/0305127.

\bibitem{mhiggsEP5} S.~Martin, 
                    {\em Phys. Rev.} {\bf D 65} (2002) 116003
                    [arXiv:hep-ph/0111209];
                    {\em Phys. Rev.} {\bf D 66} (2002) 096001
                    [arXiv:hep-ph/0206136];
                    {\em Phys. Rev.} {\bf D 67} (2003) 095012
                    [arXiv:hep-ph/0211366];
                    {\em Phys. Rev.} {\bf D 68} (2003) 075002
                    [arXiv:hep-ph/0307101]; 
                    {\em Phys. Rev.} {\bf D 70} (2004) 016005
                    [arXiv:hep-ph/0312092];
                    {\em Phys. Rev.} {\bf D 71} (2005) 016012
                    [arXiv:hep-ph/0405022];
                    {\em Phys. Rev.} {\bf D 71} (2005) 116004
                    [arXiv:hep-ph/0502168];
                    {\em Phys.\ Rev.} {\bf D 75} (2007) 055005
                    [arXiv:hep-ph/0701051];\\
                    S.~Martin and D.~Robertson,
                    {\em Comput.\ Phys.\ Commun.} {\bf 174} (2006) 133
                    [arXiv:hep-ph/0501132].

\bibitem{mhiggsFD3l} R.~Harlander, P.~Kant, L.~Mihaila and M.~Steinhauser,
                     {\em Phys.\ Rev.\ Lett.} {\bf 100} (2008) 191602
                     [{\em Phys.\ Rev.\ Lett.} {\bf 101} (2008) 039901]
                     [arXiv:0803.0672 [hep-ph]].

\bibitem{mhiggsAEC} G.~Degrassi, S.~Heinemeyer, W.~Hollik, P.~Slavich
  and G.~Weiglein, 
  {\em Eur.\ Phys.\ J.} {\bf C 28} (2003) 133
  [arXiv:hep-ph/0212020]. 

\bibitem{feynhiggs} S.~Heinemeyer, W.~Hollik and G.~Weiglein,
  {\em Comput.\ Phys.\ Commun.} {\bf 124} (2000) 76
  [arXiv:hep-ph/9812320];
  see: {\tt www.feynhiggs.de}~.

\bibitem{mhcMSSMlong}
  M.~Frank, T.~Hahn, S.~Heinemeyer, W.~Hollik, H.~Rzehak and G.~Weiglein,
  JHEP {\bf 0702} (2007) 047
  [arXiv:hep-ph/0611326].

\bibitem{chargedmhiggs} M.~Diaz and H.~Haber,
                        {\em Phys. Rev.} {\bf D 45} (1992) 4246.

\bibitem{markusPhD} M.~Frank, 
                    PhD thesis, university of Karlsruhe, 2002.

\bibitem{benchmark2} M.~Carena, S.~Heinemeyer, C.~Wagner and G.~Weiglein, 
                     {\em Eur. Phys. J.} {\bf C 26} (2003) 601
                     [arXiv:hep-ph/0202167].

\bibitem{benchmark3} M.~Carena, S.~Heinemeyer, C.~Wagner and G.~Weiglein,
                     {\em Eur.\ Phys.\ J.} {\bf C 45} (2006) 797
                     [arXiv:hep-ph/0511023].

\bibitem{cmsHiggs} S.~Gennai, S.~Heinemeyer, A.~Kalinowski, R.~Kinnunen, 
                   S.~Lehti, A.~Nikitenko and G.~Weiglein,
                   {\em Eur.\ Phys.\ J.} {\bf C 52} (2007) 383
                   [arXiv:0704.0619 [hep-ph]];
                   M.~Hashemi, S.~Heinemeyer, R.~Kinnunen, 
                   A.~Nikitenko and G.~Weiglein,
                   arXiv:0804.1228 [hep-ph].

\bibitem{Heinemeyer:2006px}
  S.~Heinemeyer, W.~Hollik, D.~Stockinger, A.~M.~Weber and G.~Weiglein,
  {\em JHEP} {\bf 0608} (2006) 052
  [arXiv:hep-ph/0604147].

\bibitem{PomssmRep} S.~Heinemeyer, W.~Hollik and G.~Weiglein, 
                    {\em Phys.\ Rept.} {\bf 425} (2006) 265
                    [arXiv:hep-ph/0412214].

\bibitem{Master1} O.~Buchmueller et al.,
                  {\em Phys.\ Lett.} {\bf B 657} (2007) 87
                  [arXiv:0707.3447 [hep-ph]].

\bibitem{Master2} O.~Buchmueller et al.,
                  {\em JHEP} {\bf 0809} (2008) 117
                  [arXiv:0808.4128 [hep-ph]].

\bibitem{Master3} O.~Buchmueller et al.,
                  {\em Eur.\ Phys.\ J.} {\bf C 64} (2009) 391
                  [arXiv:0907.5568 [hep-ph]].

\bibitem{MasterWWW} See: {\tt cern.ch/mastercode}~.

\bibitem{Allanach:2001kg} B.~Allanach,
  {\em Comput.\ Phys.\ Commun.} {\bf 143} (2002) 305
  [arXiv:hep-ph/0104145].

\bibitem{Moroi:1995yh}
  T.~Moroi,
  {\em Phys.\ Rev.} {\bf D 53} (1996) 6565
  [Erratum-ibid.\ {\bf D 56} (1997) 4424]
  [arXiv:hep-ph/9512396].

\bibitem{Degrassi:1998es}
  G.~Degrassi and G.~F.~Giudice,
  {\em Phys.\ Rev.} {\bf D 58} (1998) 053007
  [arXiv:hep-ph/9803384].

\bibitem{Heinemeyer:2003dq}
  S.~Heinemeyer, D.~Stockinger and G.~Weiglein,
  {\em Nucl.\ Phys.} {\bf B 690} (2004) 62
  [arXiv:hep-ph/0312264].

\bibitem{Heinemeyer:2004yq}
  S.~Heinemeyer, D.~Stockinger and G.~Weiglein,
  {\em Nucl.\ Phys.} {\bf B 699} (2004) 103
  [arXiv:hep-ph/0405255].

\bibitem{Heinemeyer:2007bw}
  S.~Heinemeyer, W.~Hollik, A.~M.~Weber and G.~Weiglein,
  {\em JHEP} {\bf 0804} (2008) 039
  [arXiv:0710.2972 [hep-ph]].

\bibitem{Isidori:2006pk}
  G.~Isidori and P.~Paradisi,
  {\em Phys.\ Lett.} {\bf B 639} (2006) 499
  [arXiv:hep-ph/0605012].

\bibitem{Isidori:2007jw}
  G.~Isidori, F.~Mescia, P.~Paradisi and D.~Temes,
  {\em Phys.\ Rev.} {\bf D 75} (2007) 115019
  [arXiv:hep-ph/0703035], and references therein.

\bibitem{Mahmoudi:2008tp}
  F.~Mahmoudi,
  {\em Comput.\ Phys.\ Commun.} {\bf 178} (2008) 745
  [arXiv:0710.2067 [hep-ph]]; 
  {\em Comput.\ Phys.\ Commun.} {\bf 180} (2009) 1579
  [arXiv:0808.3144 [hep-ph]].

\bibitem{Eriksson:2008cx}
  D.~Eriksson, F.~Mahmoudi and O.~Stal,
  {\em JHEP} {\bf 0811} (2008) 035
  [arXiv:0808.3551 [hep-ph]].

\bibitem{Belanger:2006is}
  G.~Belanger, F.~Boudjema, A.~Pukhov and A.~Semenov,
  {\em Comput.\ Phys.\ Commun.} {\bf 176} (2007) 367
  [arXiv:hep-ph/0607059];
  {\em Comput.\ Phys.\ Commun.} {\bf 149} (2002) 103
  [arXiv:hep-ph/0112278];
  {\em Comput.\ Phys.\ Commun.} {\bf 174} (2006) 577
  [arXiv:hep-ph/0405253].

\bibitem{Gondolo:2005we}
  P.~Gondolo, J.~Edsjo, P.~Ullio, L.~Bergstrom, M.~Schelke and E.~Baltz,
  {\em New Astron.\ Rev.} {\bf 49} (2005) 149;
  {\em JCAP} {\bf 0407} (2004) 008
  [arXiv:astro-ph/0406204].

\bibitem{Skands:2003cj}
  P.~Skands et al.,
  {\em JHEP} {\bf 0407} (2004) 036
  [arXiv:hep-ph/0311123].

\bibitem{Allanach:2008qq}
  B.~Allanach et al.,
  {\em Comput.\ Phys.\ Commun.} {\bf 180} (2009) 8
  [arXiv:0801.0045 [hep-ph]].

\bibitem{LEPSUSY} LEP Supersymmetry Working Group, see:
                  {\tt lepsusy.web.cern.ch/lepsusy/}~.

\end{susspbibliography}

\end{document}

%% file: sussp-lecture_titlepage.tex
\thispagestyle{empty}
\setcounter{page}{0}
\def\thefootnote{\fnsymbol{footnote}}

\begin{flushright}
\mbox{}
arXiv:0912.0361 [hep-ph]
\end{flushright}

\vspace{1cm}

\begin{center}

{\fontsize{15}{1} 
\sc {\bf Higgs and Electroweak Physics}}
\footnote{Lecture given at the {\em SUSSP\,65}, August 2009, St.\ Andrews, UK}

\vspace{1cm}

{\sc 
S.~Heinemeyer
\footnote{
email: Sven.Heinemeyer@cern.ch
}
}

\vspace*{1cm}

Instituto de F\'isica de Cantabria (CSIC-UC), Santander,  Spain

\end{center}

\vspace*{0.2cm}

\begin{center} {\bf Abstract} \end{center}
This lecture discusses the Higgs boson sector of the SM and the MSSM,
including their connection to electroweak precision physics and the searches
for SM and SUSY Higgs bosons at the LHC.

\def\thefootnote{\arabic{footnote}}
\setcounter{footnote}{0}

\newpage